\documentclass[12pt]{article}

\usepackage{scicite}

\usepackage{amsmath}
\usepackage{amsfonts}
\usepackage{amssymb}
\usepackage{hyperref}
\usepackage{times}
\usepackage{time}
\usepackage{graphicx}
\usepackage{color}
\usepackage[separate-uncertainty=true,per=slash]{siunitx}
\usepackage[normalem]{ulem}
\usepackage{xcolor}

 \newcommand{\beq}{\begin{equation}}
 \newcommand{\eeq}{\end{equation}}

\makeatletter
\renewcommand{\fnum@figure}{\textbf{Fig. \thefigure}}
\makeatother

\def\ie{\emph{i.e.}~}

\def\fig#1{\ref{Fig:#1}}
\def\Fig#1{Fig.~\fig{#1}}

\def\degrees{\ensuremath{^{\circ}}}

\newenvironment{sciabstract}{%
\begin{quote} \bf}
{\end{quote}}

\title{\vspace{-2.2cm} Hybrid sideways/longitudinal swimming in the monoflagellate \textit{Shewanella oneidensis}: from aerotactic band to biofilm}

\author
{Laura Stricker$^{1\ast}_\dagger$, Isabella Guido$^{2}_\dagger$, Thomas Breithaupt$^2$, Marco G. Mazza$^{2,3}$ \\
\& J\"{u}rgen Vollmer$^4$\\
\\[-2.6mm]
\small{$^{1}$ ETH Z\"urich, Department of Materials, Soft and Living Materials, 8093 Zurich, Switzerland}\\
\small{$^{2}$ Max Planck Institute for Dynamics and Self-Organization, 37077 G\"ottingen, Germany}\\
\small{$^{3}$ Loughborough University, Interdisciplinary Centre for Mathematical Modelling and}\\
\small{Department of Mathematical Sciences, Loughborough, Leicestershire LE11 3TU, UK}\\
\small{$^{4}$ University of Leipzig, Institute of Theoretical Physics, 04103 Leipzig, Germany} \\
\\[-3.4mm]
\small{$^\ast$To whom correspondence should be addressed; E-mail:  laura.stricker@mat.ethz.ch.}\\
\small{$\dagger$These authors contributed equally to the present work.}
}

\date{}

\begin{document}

\baselineskip24pt
\topmargin -1.cm
\textheight 22.cm

\maketitle


\begin{sciabstract}
\textit{Shewanella oneidensis} \mbox{MR-1} are facultative aerobic electroactive bacteria, with an appealing potential for sustainable energy production and bioremediation. They gather around air sources, forming aerotactic bands and biofilms. Though accumulation is crucial to technological exploitation, their collective behaviour remains poorly reported. Here we establish a comprehensive framework for the study of aerotaxis, unveiling a novel hybrid locomotion pattern. Despite having only one flagellum, \mbox{MR-1} combine motility features of mono- and multiflagellate bacteria, alternating longitudinal fast and sideways slow swimming. The adaptive tuning of the resulting bimodal velocity distributions fulfils different biological functions, such as aerotaxis and confinement. Overall, we reveal the mechanisms underlying the aerotactic collective behaviour of \mbox{MR-1}, in the process leading from accumulation to biofilm formation.
\end{sciabstract}


\section*{Introduction}

The energy levels of biological cells are a crucial factor for their growth and survival. Many microorganisms are able to react to a decrease in their energy levels by moving towards a microenvironment that can replenish them (energy taxis)~\cite{sch10}. Chemotaxis (energy taxis driven by chemical gradients) is ubiquitous in natural environments, regulating symbiotic interactions~\cite{rai19}. Aerotaxis is the migration of living systems towards areas with favourable oxygen concentrations for their metabolism~\cite{tay99}. Aerotactic bacteria~\cite{bar59,adl66,li08_THESIS} accumulate next to air sources forming aerotactic bands~\cite{tay99}. The strategies used by bacteria to effectively sample space in their quest for a better environment depend on their morphology~\cite{son15}. Typically, bacteria with one flagellum go fast with episodic 180\degrees~reversals or reverse-flicks~\cite{sto11}. Bacteria with multiple flagella swim slower, adopting a run-and-tumble strategy~\cite{ber72}.

In the present work, we develop a comprehensive framework for the study of aerotaxis, encompassing a novel experimental setup, numerical simulations and the analysis of bacterial trajectories. Our setup reproduces a minimalistic ecological niche, where the microorganisms modify the environment and adapt their behaviour in return. We use the Gram-negative facultative aerobic \emph{Shewanella oneidensis} \mbox{MR-1} as model bacterial strain~\cite{ven99}. These bacteria shift electrons from an electron donor towards an electron acceptor available in the environment (e.g. solid metals or oxygen) during their respiration~\cite{gor06,sub18}. Hence, they are presently considered the cornerstone for the development of sustainable technologies for energy production~\cite{log09,wu13} and wastewater treatment~\cite{rab10}, as well as biosynthesis of metal nanoparticles~\cite{ng13}, heavy metals reduction~\cite{ng13b} and biosensors~\cite{log06}.

MR-1 are extremely versatile and responsive to available resources. Their genome has been sequenced~\cite{hei02}, allowing intensive physiological and biochemical studies on their rich signal-transduction regulatory systems~\cite{fre08}. Despite the big effort put in their genetic and microbiological characterisation, their collective behaviour has not been intensively investigated. The present study fills this gap by providing a quantitative characterisation of the collective strategies adopted by \mbox{MR-1} to optimise resources, namely the formation of aerotactic bands~\cite{li08_THESIS} and air-liquid biofilms~\cite{lia10,arm13}. Understanding such strategies is the first crucial step towards technological \mbox{exploitation}.

Here we investigate the process leading from accumulation to biofilm formation and we clarify how each step is related to different motility types, in terms of velocities and directional changes. We assess the change of their locomotion strategies in response to different ambient conditions (oxygen concentration and gradient), discovering a surprisingly rich behaviour for a monotrichous bacterium. Most remarkably, besides the expected run-reverse~\cite{joh02} and run-reverse-flick patterns~\cite{xie11}, they can transition between longitudinal (fast) and sideways (slow) swimming. This mechanism allows them to access motility schemes typical of both mono- and multiflagellate bacteria. The resulting velocity distributions are bimodal and continuously changing with the local environment. 
Our work discloses how the dynamic tuning of the unique hybrid locomotion patterns of \mbox{MR-1} regulates their collective behaviour, from the accumulation in aerotactic bands to the inception of biofilm formation.

\section*{Results}

The main setup, the "closed setup" (Fig.~S1.a), consists of an air-tight closed chamber filled with \textit{Shewanella oneidensis} \mbox{MR-1} liquid bacterial culture. An entrapped micrometric air bubble provides a limited oxygen supply. The air diffuses from the bubble into the liquid, where the bacteria reduce the oxygen during their respiration, depleting it. This configuration is used to create a time-variable oxygen concentration gradient in the sample.
A second setup, the "open setup" (Fig.~S1.b), consists of a microfluidic device, partially filled with the bacterial culture, open to the air on one side and closed on the other side. The air side acts as an unlimited oxygen supply. This configuration is used to create a time-constant oxygen concentration gradient.
Imaging is done with dark-field and phase-contrast microscopy.
\\[-7mm]
\subsubsection*{Dynamic evolution of the aerotactic band}
\label{sec:Observed phenomenon}

In the closed setup, \mbox{MR-1} form an aerotactic band around the bubble (Movie~S1, \Fig{Band formation, experiment}). We describe here the formation and evolution of the band.

Initially, the bacteria have a uniform concentration and a high motility in the whole sample.
After 20~minutes, they start to aggregate around the bubble, forming a distinct band (\Fig{Band formation, experiment}.a-b). The bacteria move inside the band and in the area between the bubble and the band, while the rest of the sample enters a non-motile vibrational state (Movie~S2). As the bacteria consume the oxygen, the band progressively advances towards the bubble (\Fig{Band formation, experiment}.c), leaving behind a depletion layer (\Fig{Band formation, experiment}.d) with non-motile vibrating bacteria. Eventually, the band reaches the bubble (\Fig{Band formation, experiment}.e) and disappears (\Fig{Band formation, experiment}.f). At this point, all the bacteria are in the non-motile vibrational state. Depending on initial conditions, the band forms in 5 to 40~minutes and disappears 15 to 60~minutes after closing the sample.

We provide a quantitative evaluation of the phenomenon, by tracking the time evolution of the concentration
profiles of \textit{Shewanella} around the bubble (\Fig{Shewanella profiles}).
The bacterial density is estimated from dark-field images, as proportional to the local light intensity.
We compare the experimental results (solid lines) with the predictions from an in-house numerical model (dashed lines) (Supplementary Text),
finding good qualitative and quantitative agreement. The model captures how the maximum bacterial concentration increases, as the band moves towards the bubble. The time scales for the band formation and disappearance are reproduced within a 5\% and 15\% error, respectively.

In the open setup, the observed phenomenon is analogous but a steady state is reached after 2 hours: the \textit{Shewanella} concentration profile becomes stable, with a band at a fixed distance from the air-liquid interface (Fig.~S2.a). The bacteria in the band retain their motility and never enter the vibrational state. The observation is unaltered after 24~hours.
We conclude that the transition between motile and non-motile bacteria observed in the closed setup corresponds to the transition between aerobic and anaerobic functioning.
\\[-7mm]
\subsubsection*{The oxygen concentration regulates the biofilm formation}
\label{sec:Air-liquid interface}

\textit{Shewanella} form a biofilm called pellicle at air-liquid interfaces~\cite{lia10,arm13}. We clarify here how the oxygen level regulates the process.

\begin{figure}
\centering
\includegraphics[width=0.7\textwidth]{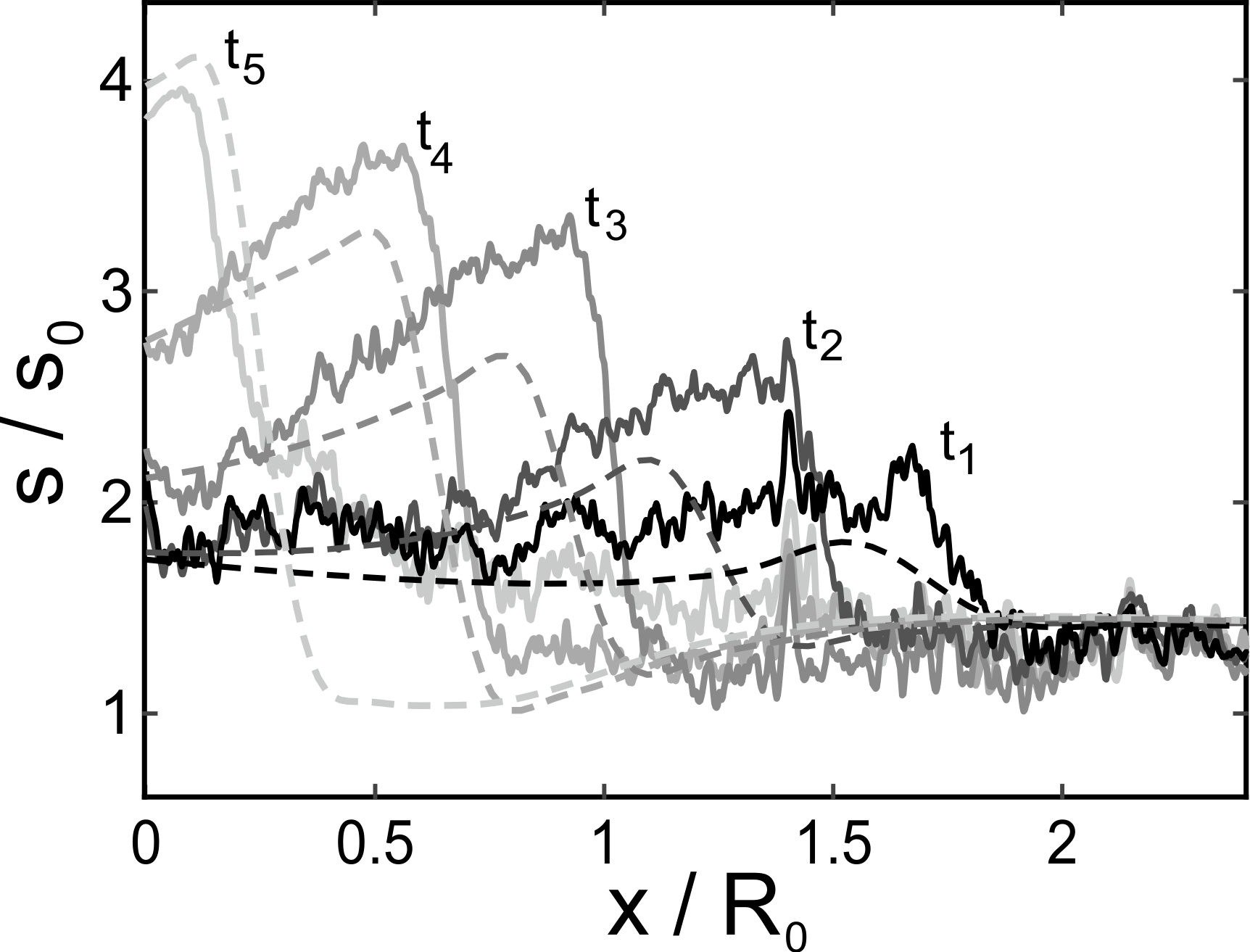}
\caption{\textbf{Time evolution of bacterial concentration profiles around a confined bubble.} Solid lines depict experiments in the closed setup, dashed lines simulations; $x$ is the distance from the interface, $s$ the bacterial concentration, \mbox{$R_0 = 193\,\micro$m} the initial bubble radius and \mbox{$s_0 = 7.11\cdot 10^7$~bacteria/ml} the initial bacterial concentration. The grey scale represents different time instants, increasing from darker to lighter: \mbox{$t_1$ = 18 min}, \mbox{$t_2$ = 21 min}, \mbox{$t_3$ = 23 min}, \mbox{$t_4$ = 24 min}, \mbox{$t_5$ = 25 min}.}
 \label{Fig:Shewanella profiles}
\end{figure}

With dark-field imaging, we do not have direct access to the air-liquid interface itself, which is saturated by the high amount of scattered light.
However, we can extract equivalent information by imaging the bacterial concentration in the liquid in its immediate neighbourhood.
In the closed setup, such a concentration remains constant during the initial phases of the band formation
(\Fig{Shewanella profiles}, curves $t_1$, $t_2$); later on, it increases while the band moves towards the interface
(\Fig{Shewanella profiles}, curves $t_3-t_5$).
This behaviour indicates that, initially, the bacterial flux at the bubble wall is entirely adsorbed, while it gets progressively reflected as the oxygen supply decreases.
Conversely, in the open setup, where the oxygen supply is unlimited and the oxygen concentration
at the interface is constant, the nearby bacterial concentration remains unvaried throughout the whole process. This observation indicates that
the incoming flux is entirely adsorbed by the interface.

\begin{figure}
\centering
\includegraphics[width=0.7\textwidth]{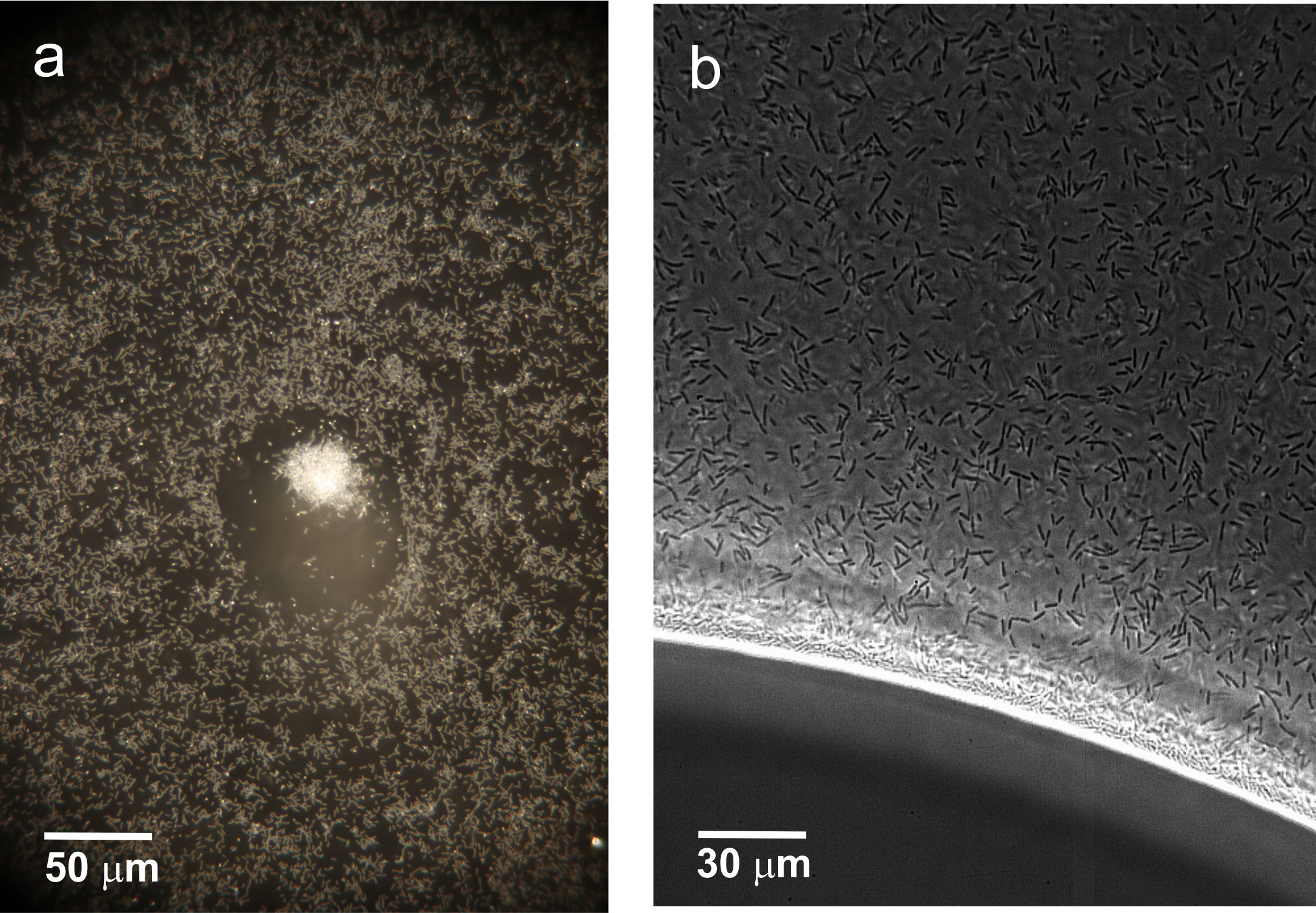}
\caption{\textbf{Biofilm precursor formed at the air-liquid interface, with limited oxygen supply.} The closed setup is imaged after the band has reached the bubble; (a) dark-field imaging shows a depletion layer, around the bubble, with a reduced bacterial density; (b) phase-contrast imaging reveals a nematic-like layer of piled up bacteria at the bubble interface.}
 \label{Fig:Interface air-liquid}
\end{figure}

Phase-contrast microscopy reveals a layer of bacteria with nematic-like ordering, piled up at the air-liquid interface and surrounded by a depletion zone. In the closed setup, entailing a limited oxygen supply, such a layer does not increase indefinitely, but stops growing presumably below a certain oxygen concentration (\Fig{Interface air-liquid}). Conversely, with an unlimited oxygen supply, the layer keeps growing (Fig.~S3). We conclude that the oxygen concentration regulates the adsorption of \mbox{MR-1} at the air-liquid interface.

On occasion, small droplets are captured inside the bubble in the closed setup. Their observation allows direct imaging of the pellicle formation process (Fig.~S4, Movie~S3).
Here, two air-liquid interfaces are present: (i) between the bubble and the liquid,
where a moving aerotactic band appears, and (ii) between the droplet and the bubble.
On the latter interface, we observe the presence of bacterial clusters with an active motion. They progressively grow in size, by incorporating colliding swimming bacteria (Movie~S4).
The process slows down in time. Eventually, the bacteria stop moving and the clusters stop growing. This happens exactly when the band surrounding (i) reaches the bubble, \ie when the lack of oxygen induces the anaerobic transition everywhere. We infer a connection between bacterial motility and pellicle formation, both regulated by the oxygen level. This conclusion is in line with previous findings that a fully functional flagellum is a prerequisite for pellicle formation in \mbox{MR-1}~\cite{arm13}.
\\[-7mm]
\subsubsection*{Hybrid locomotion patterns: alternated sideways/longitudinal swimming}
\label{sec:PDF and tracking}
We show here that \mbox{MR-1} feature hybrid locomotion patterns, even in the absence of oxygen gradients. The trajectories of the motile bacteria are analysed in a variation of the closed setup where no bubble is present, immediately after closure of the sample (\ie with high uniform oxygen concentration).
\\[3.7mm] \indent
To evaluate the motility, we consider the probability distribution function (PDF) of the absolute values of instantaneous velocities, derived from the time sampling of trajectories.
Around 15\% of bacteria perform a stop-and-go type of motion: they alternate short pauses and runs (Movie~S5). Hence, there is a peak in the PDF, around 0~{\micro}m/s (Fig.~S5). We will show in the next section that this is consistent with the inception of biofilm formation. By removing the pauses and accounting only for the motile parts of trajectories, the PDF becomes bimodal, with peaks around 25~{\micro}m/s and 75~{\micro}m/s (\Fig{PDF velocities, no gradient}.a). A priori, this could correspond to two different scenarios: two populations of bacteria with a constant velocity (fast or slow), or bacteria changing their velocity, by alternating fast and slow runs. We compare the PDF of the instantaneous velocities with the PDF of the average velocities along the trajectories (Fig.~S6). Coincidence between the two PDFs would indicate that there are two populations with constant velocity and no switching. We observe that the PDF of the average velocities along the trajectories is still bimodal but the relative amplitude of the peaks has changed. Therefore, we conclude that both scenarios occur: there are fast and slow bacteria but also bacteria changing their velocity along a single trajectory, as confirmed by visual inspection.
\\[3.5mm]
\indent
To assess how the speed is related to the reorientation strategies, we analyse the angles between consecutive portions of trajectories and we sort them based on the average speed preceding the turn (\Fig{PDF consecutive angles, no gradient}). We find three regimes with significant differences: slow (below 25~{\micro}m/s), fast (above 60~{\micro}m/s) and intermediate (25-60~{\micro}m/s). Slow bacteria span all angles between 0\degrees~and 180\degrees; more than 80\% of such bacteria swim sideways with respect to their main body axis. Fast bacteria swim with straight or continuously curving trajectories, parallel to their longitudinal axis, and turn by small angles (up to 70\degrees). Bacteria with intermediate velocities exhibit an intermediate behaviour, performing either small angles (0\degrees-80\degrees) or reversals (140\degrees-180\degrees); they mostly swim parallel to their main axis but can switch to sideways motion (Movie~S6).
\\[3.5mm]
\indent
This differentiation within the same population of monoflagellate bacteria is remarkable, as it combines features of typical strategies used by morphologically different species to explore the surroundings. Monotrichous bacteria, such as $\sim$70\% of marine bacteria, are typically fast (up to 75~{\micro}m/s), moving by straight or curved trajectories~\cite{joh02}. They change direction by inverting their flagellar rotation, inducing reversals (150\degrees-180\degrees~angles), like \mbox{MR-1} in the intermediate velocity runs. Peritrichous bacteria, such as most enteric bacteria, are typically slower ($\sim$30~{\micro}m/s) and move by run-and-tumbling~\cite{joh02}. They swim in almost straight runs, with the flagella bundled together, stopping and tumbling, when one flagellum inverts its rotation disrupting the bundle. All angles occur upon reorientation, like in the slow runs of \mbox{MR-1}~\cite{ber72}.
\\[3.5mm]
\indent
Hybrid locomotion mechanisms have been reported for several bacterial species, as a way to enhance direction randomisation~\cite{sto11}.
Other monotrichous bacteria alternate 180\degrees~reversals with 90\degrees~flicks induced by the flagellum bending (run-reverse-flick)~\cite{xie11,sto11}. Though \mbox{MR-1} can perform this type of motion, the absence of the peak at 90$\degrees$~shows that it is not their main strategy. The predominance of sideways swimming in the slow regime, together with the possibility to switch between sideways and longitudinal motion, suggests that they generate torque at low velocities by modulating the coiling shape of the flagellum. This is in line with the recent discovery that some bacteria, including the conspecific \textit{S.~putrefaciens}, can swim with the flagellum wrapped around their body~\cite{hin17,kuh17}. 
\\[-7mm]
\subsubsection*{The adaptive tuning of sideways/longitudinal swimming shapes the collective behaviour}
We study here how the ambient conditions influence the motility and the reorientation strategies of \mbox{MR-1}, determining different collective behaviours, namely band and biofilm formation. We perform two experiments to separately test the impact of the oxygen concentration and the oxygen gradient. To this aim, we use two variations of the closed setup: without bubble (\ie uniform oxygen concentration, decaying in time) and with a bubble (\ie with an oxygen gradient).
\\[2.mm]
\indent To evaluate the effect of the oxygen concentration, we examine the samples without bubbles for 15~minutes after sealing. As the bacteria deplete the oxygen, the peak at slow velocities in the PDF of instantaneous velocities progressively disappears (\Fig{PDF velocities, no gradient}.b). The PDF becomes unimodal with a peak around 80~{\micro}m/s. This observation is compatible with the interruption of the biofilm formation at low oxygen concentration. We conclude that, like \textit{S. putrefaciens}~\cite{bar03}, \textit{S. oneidensis} adapts its speed in response to the local concentration of a chemoattractant. Hence, it does not only perform chemotaxis (reaction to a concentration gradient), but also chemokinesis (reaction to the concentration itself).
\\[3.5mm]
\indent To investigate the effect of the oxygen gradient, we analyse the case with an entrapped bubble, featuring a moving band.
\Fig{PDF velocities, different points} displays the PDF of instantaneous velocities, at three different locations: (a)~at the air-liquid interface, (b)~between the bubble and the band ('middle') and (c)~on the band.
The numerical simulations show that the oxygen concentration is always monotonically decreasing in the radial direction. At the interface and in the middle, the PDFs reproduce the trend observed without bubbles, for decreasing oxygen concentrations: the bimodal distribution in (a) turns unimodal in (b), where the oxygen concentration is lower. On the band, however, such a trend is disrupted: the peak at slow velocities reappears and the number of particles with an intermediate speed (25-60~{\micro}m/s) increases. The PDFs of turning angles, sorted by velocity ranges, are unvaried with respect to the case without oxygen gradient.
We conclude that different speed ranges correspond to different reorientation strategies, fulfilling specific biological functions. In particular, a peak at slow velocities, where \textit{Shewanella} reorients in any direction, emerges in environmental conditions where there is no preferential direction of swimming. At the bubble wall, it relates to the inception of the biofilm formation. In the areas where the bacteria accumulate (aerotactic band), it reflects the fact that \mbox{MR-1} is in its optimal oxygen concentration range. On the band, the increased number of bacteria swimming at intermediate velocities (25-60~{\micro}m/s), favouring straight runs or reversals, guarantees the persistence in the region: when a bacterium exits the band, it either stops or immediately comes back.
In the region between the bubble and the band, high velocities are dominant (above 60~{\micro}m/s), mostly associated to straight runs. This is typical of directional motion and reflects the aerotaxis towards the band.
Summarising, slow velocities are associated to sideways displacements and isotropic direction randomisation, intermediate velocities present run-reverse patterns producing confinement, high velocities encompass straight runs, related to directional aerotactic motion.

\section*{Conclusions}
The present work is a comprehensive study of the collective behaviour of \textit{Shewanella oneidensis} \mbox{MR-1} next to an air source, from accumulation to the air-liquid biofilm formation. We studied the dynamic evolution of the aerotactic band gathering around a bubble, both with experiments and simulations. When the air source is not confined, the band remains at a constant distance from the air-liquid interface. Conversely, when the air source is confined and the oxygen is depleted by the bacteria, the band moves towards the interface, eventually reaching it and disappearing. The local oxygen concentration regulates the biofilm formation, by affecting the bacterial adsorption at the interface and their motility. Both are suppressed when the oxygen is entirely depleted.
\\[3.3mm] \indent
The system at hand is a minimal model of an ecological niche, where the inhabitants modify the environment and their behaviour dynamically adapts in return.
We characterised the collective locomotion strategies of \mbox{MR-1} in response to the environmental conditions (oxygen concentration and gradient), by tracking their trajectories. They explore the surroundings with a combination of fast and slow runs, spanning a broad range of velocities. Their velocity distributions are bimodal and dynamically change with the ambient conditions, posing a challenge to the concept of average velocity of a bacterial population. Different velocities are associated to specific reorientation strategies, realising different biological functions, such as aerotaxis or confinement. Fast runs are mostly straight or curved; their number increases when the bacteria perform aerotaxis in response to an oxygen gradient.
Slow runs correspond to a higher direction randomisation, with turns spanning all angles. They are associated to favourable conditions for the bacteria, such as their preferred oxygen concentration range in the aerotactic band, or to the inception of the biofilm formation. Surprisingly for a monotrichous bacterium, they are mostly realised through sideways swimming. At intermediate velocities the bacteria alternate straight runs and reversals. In the aerotactic band this behaviour supports confinement, correcting the trajectories leading away from the optimal oxygen concentration.
The possibility to switch between fast (longitudinal) and slow (mostly sideways) swimming allows \mbox{MR-1} to access typical velocity and reorientation ranges of both mono- and multiflagellate bacteria, tremendously increasing their biological competitiveness.
The present work establishes the link between the intriguing hybrid locomotion strategies of \mbox{MR-1} and their aerotactic collective behaviour, showing how the adaptive tuning of their bimodal velocity distributions regulates the process leading from confinement in aerotactic bands to biofilm formation.

\clearpage


\clearpage

\begin{figure}
\centering
\includegraphics[width=0.7\textwidth]{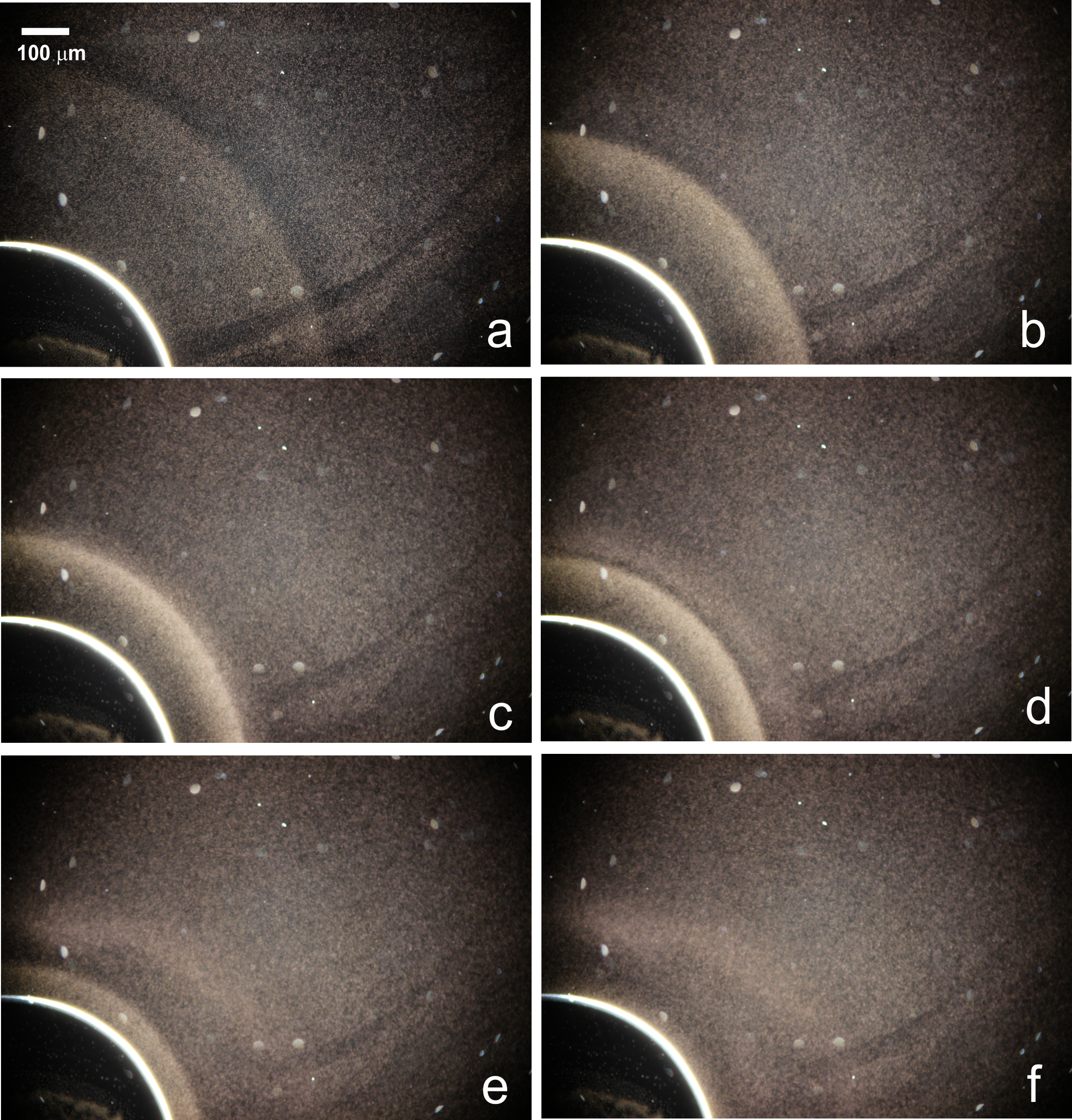}
\caption{\textbf{Time evolution of the aerotactic band around a confined bubble.} The initial bubble radius is \mbox{$R_0 = 328\,\micro$m}. The closed setup is imaged in dark-field (a)~20~min, (b)~25~min, (c)~30~min, (d)~37~min, (e)~42~min, (f)~50~min after sealing the sample.}
 \label{Fig:Band formation, experiment}
\end{figure}

\begin{figure}
\centering
\includegraphics[width=0.7\textwidth]{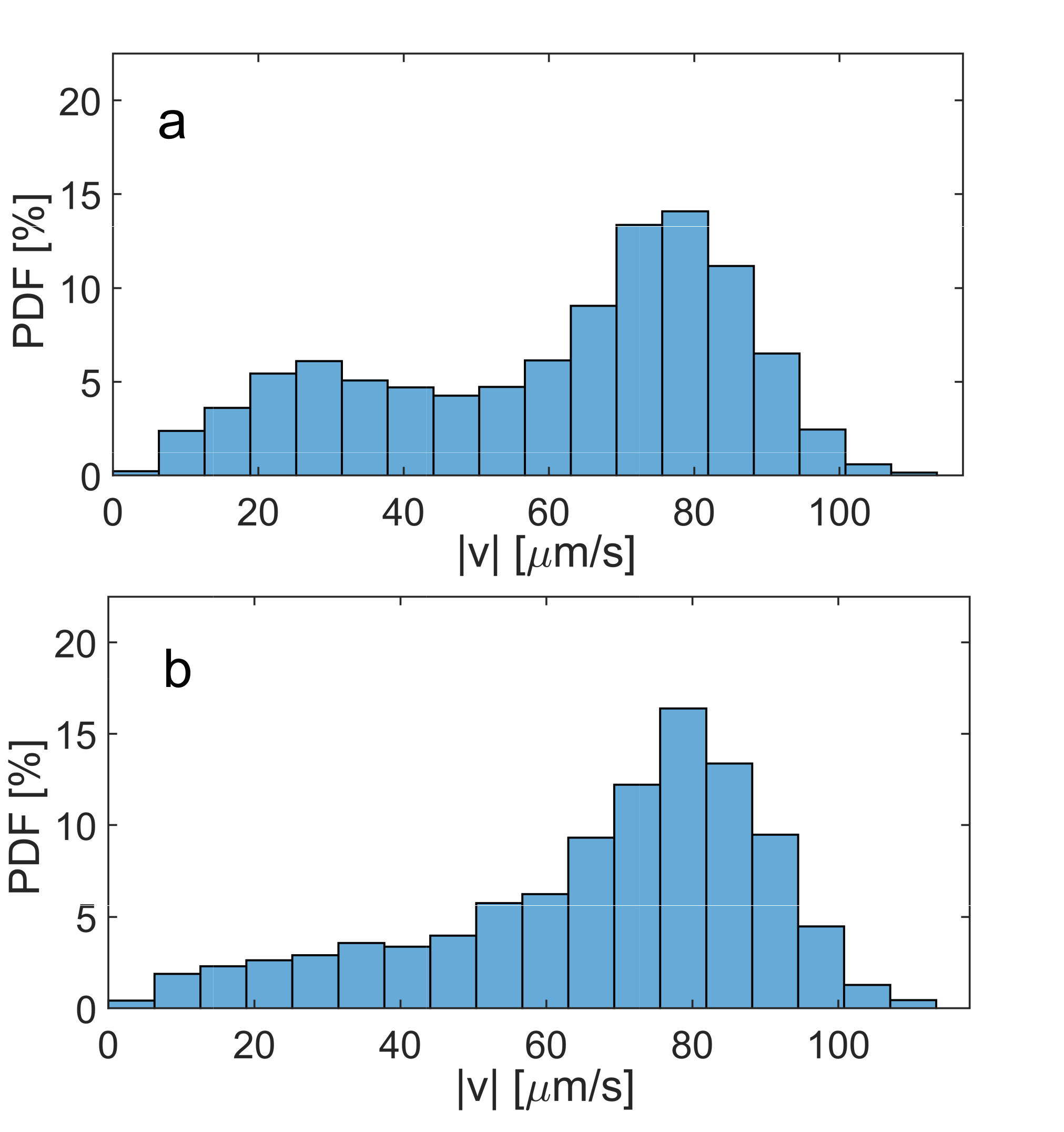}
\caption{\textbf{PDF of instantaneous velocities of \mbox{MR-1} for different oxygen concentrations.} The closed setup without bubble is imaged (a)~1~min and (b)~10~min after sealing the sample. Each histogram includes $\sim$~4,000 points.}
 \label{Fig:PDF velocities, no gradient}
\end{figure}

\begin{figure}
\centering
\includegraphics[width=0.8\textwidth]{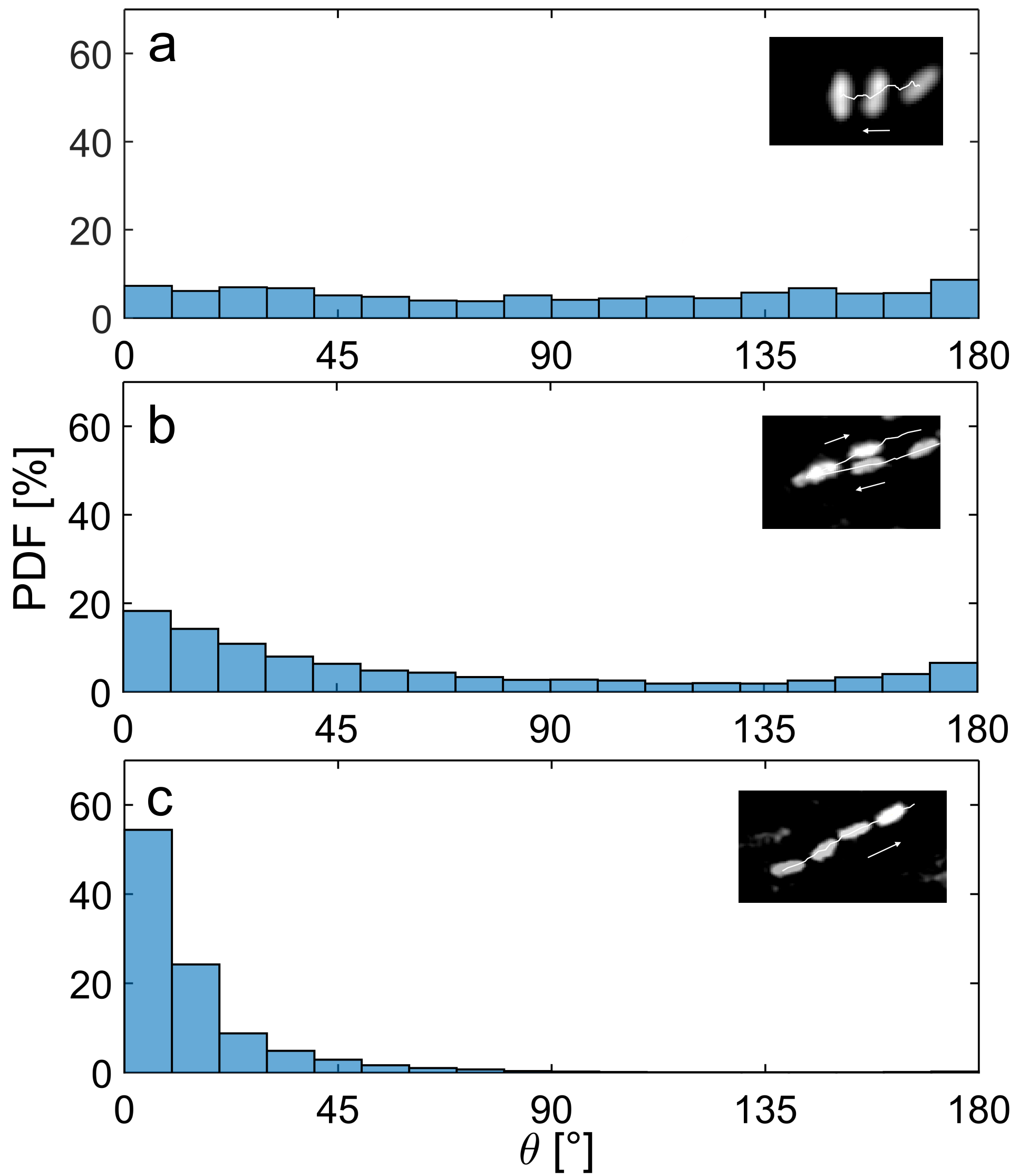}
\caption{\textbf{PDF of consecutive angles performed by MR-1, for different instantaneous velocity ranges.} (a)~Below 25~{\micro}m/s, (b)~between 25~{\micro}m/s and 60~{\micro}m/s and (c)~above 60~{\micro}m/s. The inserts depict typical respective locomotion patterns of the bacteria: (a) sideways, (b) run-reverse and (c) straight longitudinal runs. They represent the stroboscopic mapping of dark-field images with constant time lags of (a)~0.1 s, (b)~ 0.06 s, (c)~0.04 s. The arrows indicate the direction of motion.}
 \label{Fig:PDF consecutive angles, no gradient}
\end{figure}


\begin{figure}
\centering
\includegraphics[width=0.8\textwidth]{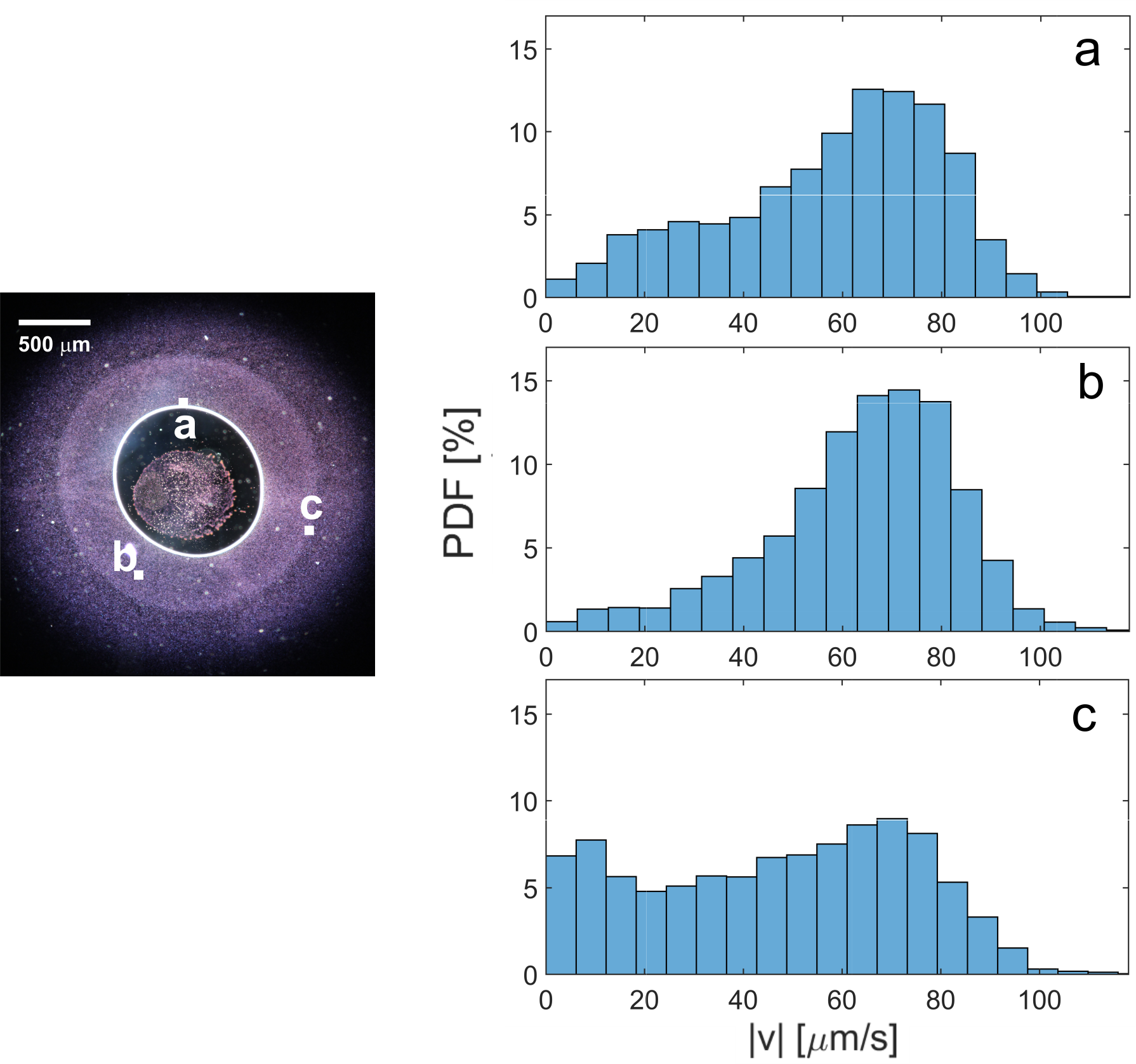}
\caption{\textbf{PDF of the instantaneous velocities of MR-1 at different locations, around a bubble.} The initial bubble radius is $R_0 = 514\, \micro$m. The closed setup is imaged 20 minutes after sealing the sample: (a)~at the bubble interface, (b)~between the bubble and the aerotactic band and (c)~on the band.  
Each histogram includes $\sim$~10,000 points.}
 \label{Fig:PDF velocities, different points}
\end{figure}

\clearpage

\topmargin -1.5cm
\textheight 24.cm

\section*{Acknowledgements}
We thank Laura Turco for technical support in the experiments, Gal Scholnik for performing preliminary experiments,
Andreas Kappler for providing the bacterium \textit{S. oneidensis} \mbox{MR-1} and Mathias Schr\"{o}ter for support in preliminary bacterial tracking.
We acknowledge discussions with Anupam Sengupta and Martin Kr\"{o}ger.
We thank Hans-Christian \"{O}ttinger, James Clewett and Martin Callies for feedback on the manuscript.
\textbf{Funding} We acknowledge financial support from the Deutsche Forschungsgemeinschaft (SFB 937, project A20) and from the People Programme (Marie Curie Actions) of the European
Union's Seventh Framework Programme FP7/2007-2013/ under REA grant agreement n\degrees [628154].
\textbf{Author Contributions}
I.G., L.S. and M.M. designed the experiments.
I.G. and L.S. performed the experiments.
L.S. and J.V. developed the numerical model.
T.B. and L.S. did the tracking.
L.S., I.G. and J.V. did the statistical analysis and interpretation of the data.
J.V. and M.M. contributed with comments.
L.S., I.G. and J.V. wrote the draft. All the authors contributed to subsequent revisions.
\textbf{Competing Interests} The authors declare that they have no competing financial interests.
\textbf{Data and materials availability}
All data needed to evaluate the conclusions in the paper are present in the paper and/or the Supplementary Materials.
Any additional data supporting the findings of this study as well as the in-house codes are available from the corresponding author upon request.
Correspondence and requests for materials should be addressed to L.S (email: laura.stricker@mat.ethz.ch).
\\[-10.mm]
\section*{Supplementary materials}
Materials and Methods\\
Supplementary Text\\
Supplementary References\\
Fig.~S1-S6\\
Movie~S1-S6.

\end{document}


\maketitle

\baselineskip24pt
\topmargin -1.cm
\textheight 22.cm

\section*{Materials and methods}

\subsection*{Cultivation}
We use \textit{Shewanella oneidensis} \mbox{MR-1} bacteria (Zentrum f\"{u}r Angewandte Geowissenschaften, Universit\"{a}t T\"{u}bingen). The bacteria are cultivated in aerobic incubation with 150~rpm shaking, first in Luria-Bertrani broth (Roth, Karlsruhe, Germany), then inside minimal medium (MM) \cite{mel14} with 20~mM sodium lactate (Roth, Karlsruhe, Germany) as substrate, following the procedure described in \cite{sch16}. The latter culture is harvested in the early exponential growth 
and used for the experiments. All the figures are realised using cultures with $OD_{600} = 0.3$, unless otherwise specified.

\subsection*{Experimental setup}
We have two types of setup, featuring an air-tight closed chamber ("closed setup") and a chamber open to the air ("open setup"). \\
- \textbf{Closed setup}
We build the closed chamber using a microscope glass slide, a coverslip and a spacer, with the following procedure. We prepare the glass slide and the coverslip by washing them in consecutive steps in propanol, water and acetone, to prevent adhesion of the bacterial flagellum to the surface. By means of a double-coated adhesive tape 30, 50 or 100~{\micro}m thick (No.~5603, No.~5605 and No.~5015P, Nitto Denk Corporation, Japan), we fabricate a mask \mbox{40$\times$40~mm} with an internal square-shaped chamber \mbox{15$\times$15~mm} and we attach it to the glass slide \mbox{76$\times$52$\times$1~mm} (Marienfeld, Germany). Using a pipette, we deposit few droplets of the bacterial culture inside the internal chamber. We then cover the chamber with a coverslip \mbox{18$\times$18~mm} (No.~1.5, Menzel-Gl\"{a}ser, Germany) and uniformly distribute a thick layer of silicon vacuum grease  (Dow Corning, Midland, MI, USA) all around the mask and the coverslip, to prevent air leakages into the chamber. One or more bubbles remain trapped inside the chamber. We discard the samples with bubbles with eccentricity higher than $0.5$ or closer than 10 bubble radii to each other or to the chamber walls. To produce samples without air bubbles, for the control experiments and the tracking of trajectories, the coverslip is deposited in a slower fashion.\\
- \textbf{Open setup}
We build the open chamber using the microfluidic device $\mu$-Slide I Luer (Ibidi, Martinsried, Germany), in combination with a glass coverslip D~263~M Schott (\mbox{2.5$\times$7.5~cm}). After assembly, the device consists of a rectangular channel \mbox{50~mm$\times$5~mm$\times$450~{\micro}m}, with the extremes connected to the air by means of two outlets. The culture medium with the bacteria is introduced inside the channel with a syringe, creating a meniscus, located approximately in the middle of the channel. The outlet on the liquid side is closed with a plastic cap, and made air-tight by applying a thick layer of silicon vacuum grease (Dow Corning, Midland, MI, USA). The other outlet, on the air side, is left open.

\subsection*{Imaging}
For dark-field imaging, we use an upright microscope Leica DM 2500M (10$\times$, 20$\times$ and 40$\times$ magnification).
For phase-contrast imaging, we use an inverted microscope Leica DMi8 (100$\times$ magnification).
The image acquisition of the pictures is done with a Canon EOS 600D camera.
The videos for the particle tracking are taken with a sCMOS Camera pco.edge 4.2 (PCO AG, Germany) at 100~fps (20$\times$ and 40$\times$ magnification).

\subsection*{Particle tracking}
We analyse the trajectories of the bacteria, with the 2D particle tracking Matlab software u-Track \cite{jaq08}. We individually check the tracks by visual inspection to correct for occasional mistakes and to determine the orientation of the body axes respect to the direction of motion.
The videos are pre-processed with an in-house code, developed in Matlab, to remove the sessile bacteria and the occasional light reflections in the background. The results of the tracking are \mbox{analysed} with another in-house software, written in Matlab. For the details of the pre- and post-processing algorithms and the optimal parameters used for the tracking, we refer the reader to \cite{bre18}. Examples of the videos used for the analysis of trajectories can be found in the Supplementary Movies.

\subsection*{Statistical analysis}
To calculate the probability distribution functions (PDF) of the instantaneous velocities and turning angles, we sample the trajectories using either constant time or constant length intervals. The two methods are equivalent to applying a time or a space filter, respectively. Too small intervals mostly display the noise, too big intervals select bigger scale phenomena, but encompass a smaller number of points, hence a deterioration of the statistics. Appropriate intervals are \mbox{$7 \cdot 10^{-2}$~s} for time sampling and $3$ bacterial lengths (\mbox{$\sim 10$~{\micro}m}) for space sampling. We verify that the time and space sampling methods give equivalent results.

\subsection*{Control experiment with passivated bacteria}
We verify that the formation of the band is due to the active swimming of \mbox{MR-1}. In particular, we rule out evaporation and the consequent coffee stain effect \cite{dee97} as a reason for the bacterial accumulation at the air-liquid interface.
To this end, we suppress the active motility of the bacteria, without changing their shape, by heating them at 65$\celsius$ for 60~minutes. We inoculate the passivated bacteria in a closed setup with a bubble and we check by visual inspection that their active motility is indeed suppressed. We image the behaviour of the bacteria every 10~minutes, for 2~hours. The concentration remains uniform; hence, the bacterial accumulation around the bubble is due to their active motility.

\subsection*{Control experiment for the role of light}
We verify that the bacteria are attracted by the oxygen inside the bubbles and not by the scattered light.
To this end, we prepare a closed setup without bubbles and otherwise identical conditions as those adopted to study the band formation. We apply a light spot by positioning the sample under a microscope Olympus 81 and shedding the microscope built-in light through its objective. This results into a localised beam with a radius of \mbox{150~{\micro}m}. We image the behaviour of the bacteria for 2~hours, by switching to dark-field every 10~minutes. The bacterial concentration remains uniform. Hence, the bacteria are not attracted by the light source.

\subsection*{Reproducibility}
The aerotactic bands described in the present work have been observed over a sample of more than 50~repetitions.
The droplets entrapped inside the bubble have been observed 4~times. Such samples were discarded from the quantitative analysis, but still observed, as they provided anecdotal evidence of the pellicle formation process.

\subsection*{Bacterial concentration profiles from experiments}
We derive the bacterial concentration as a function of the distance from the bubble, from the dark-field images. We consider the local bacterial concentration to be proportional to the local light intensity. The luminosity profiles are extracted with the software Fiji. The bubble interface itself is saturated hence inaccessible. However, the saturated area has a constant thickness, depending on the adopted magnification (e.g. \mbox{12$\pm$2~{\micro}m} with 40$\times$ magnification). The area in the immediate proximity of the boundary also has a higher luminosity, due to the light reflection from the interface, but it is not saturated. To account for this effect, we preliminary measure the luminosity field surrounding a bubble in the absence of bacteria, on a sample of 10~bubble sizes. Fitting these data with a decaying exponential function provides the decay length for each bubble radius. By interpolating the values measured for the different radii, we derive the radius dependence of the decay length on the bubble radius. The corresponding luminosity profile is then subtracted from the experimental images with the bacteria.

\subsection*{Model and equations}
We develop a model based on a continuum field description, consisting of a system of coupled differential equations. We consider an idealised axially-symmetric geometry, with a cylindrical air bubble immersed in an infinite liquid medium. The gas inside the bubble (air) is treated as a binary mixture of perfect gases, \ie nitrogen and oxygen, diffusing into the liquid, causing the bubble to shrink. The oxygen is depleted by the bacteria. The nitrogen is only passively diffusing and does not interact with the bacteria, but it is included to prevent the total dissolution of the bubble. 
The model describes the evolution of the bubble radius $R(t)$ and the concentration fields of oxygen $c(r,t)$, nitrogen $n(r,t)$ and \textit{Shewanella} $s(r,t)$, where $t$ is the time and $r$ the radial coordinate.
The flux of bacteria is modelled as the superposition of a diffusive flux, due to their random motility, and a chemotactic flux, responsible for the drift along the oxygen gradient~\cite{kel71,kel71b}. The bacterial growth and the transition between aerobic and anaerobic state are kept into account; the bacterial death is neglected, as it is irrelevant on the time scales of the experiment. The mechanism described in the section 'Results. The oxygen concentration regulates the biofilm formation' for the partial adsorption of \mbox{MR-1} at the air-liquid interface is incorporated in the boundary conditions.
The temperature of the system $T_{\infty}$ is homogeneous and constant in time. We summarise here the equations of the model and we refer the reader to 'Model derivation' and 'Estimate of parameters' in the Supplementary Text for further details.\\
- \textbf{Time-evolution of the bubble radius}
\beq \dot{R} = \frac{T_{\infty}}{ p_g - \frac{\sigma}{2R} } \biggl[ D_{O_2} \mathcal{R}^*_{O_2} \pder{c}{r}\biggr|_{R(t)} + D_{N_2} \mathcal{R}^*_{O_2} \pder{n}{r} \biggr|_{R(t)} \biggr] \, ,\label{Eq:dR/dt}\eeq
where the dot denotes a time derivative, $p_g$ is the total pressure inside the bubble, $\sigma$ the surface tension, $D_{O_2}$, $D_{N_2}$ the diffusive constants and  $\mathcal{R}^*_{O_2}$, $\mathcal{R}^*_{N_2}$ the specific gas constants of oxygen and nitrogen.\\
- \textbf{Advection-diffusion equation for the nitrogen concentration} \cite{str11,shp13b}
\beq \label{Eq:gas concentrat eqn, N2} \pder{n}{t}+ \frac{R\dot{R}}{r}\pder{n}{r}= D_{N_2} \nabla^2 n \, . \eeq
- \textbf{Advection-diffusion equation with consumption for the oxygen concentration}
\beq \label{Eq:gas concentrat eqn, O2} \pder{c}{t} + \frac{R\dot{R}}{r}\pder{c}{r} =  D_{O_2} \nabla^2c - A_0 s\frac{c}{C_s+c}\, , \eeq
where $A_0$ is a constant indicating the mass of oxygen consumed per mass of \textit{Shewanella} produced, and
$C_s$ is the half-saturation constant of \textit{Shewanella}. \\
- \textbf{Advection-diffusion-aerotaxis equation for the \textit{Shewanella} concentration}
\beq \pder{s}{t} + \frac{R\dot{R}}{r}\pder{s}{r} = \nabla \cdot [\mu(c) \nabla s - \chi_0 s \alpha(c) \nabla c] +  \nu \frac{c}{C_s + c} \, . \eeq
The terms under divergence denote the bacterial flux, $\nu$ is the maximum specific growth rate, $\chi_0$ the chemotactic sensitivity, $\mu(c) = \mu_0 \text{H}(c-c_T) $ is the random motility coefficient, with $\mu_0$ its maximum value, $\text{H}$ the Heaviside function and $c_T$ the oxygen concentration corresponding to the aerobic/anaerobic transition; $\alpha(c)$ is a function expressing the dependence of the aerotactic response on the local oxygen concentration
\beq
\alpha(c) = \frac{K_D}{(K_D + c)^2} \text{H}(c-c_T) [1 - \text{H}(c-c_{opt})]  \, ,
\label{Eq:alpha(c) heuristic}
\eeq
with $K_D$ the receptor dissociation constant, $c_{opt}$ the optimal oxygen concentration for the bacteria, $\text{H}(c-c_T)$ and $\text{H}(c-c_T)$ the Heaviside functions centred around $c_T$ and $c_{opt}$ respectively, replaced by their smeared versions in the numerical treatment \cite{xu12}. Our heuristic formulation is in line with the classical formulations \cite{lap76,riv89}, accounting for down-regulation (the effect of attractant concentration on the number of expressed cell surface receptors) and receptor saturation, but it additionally incorporates the suppression of aerotaxis when the bacteria reach their favourite concentration range, and the transition to anaerobic functioning at low oxygen concentration (Fig.~S5).
For the boundary conditions at infinity, we take all concentration fields with null spatial derivative. At the bubble wall, for oxygen and nitrogen, we assume equilibrium with the gas inside the bubble, by means of Henry's law; for \textit{Shewanella} we assume partial adsorption, regulated by the oxygen, in line with the experimental findings (see 'Results. The oxygen concentration regulates the biofilm formation'). 
Hence, we impose the bacterial flux $J$ at the bubble wall
\beq
J|_{R(t)} = r_as\epsilon(c)|_{R(t)} \, ,
\label{Eq:BC Shewanella at R(t)}
\eeq
with $r_a$ the adsorption constant and $\epsilon(c)$ an increasing function of the oxygen concentration; for simplicity $\epsilon(c) = c$.

\subsection*{Numerical method}
The system is solved with a pseudospectral collocation method, previously developed and tested in \cite{str11,str12_thesis}, implemented with an in-house code written in Fortran. Details on the implementation are given in 'Numerical method' in the Supplementary Text.

\clearpage

\section*{Supplementary text}

\subsection*{Model derivation}
\label{sec:Model derivation}
In the experiments, an air bubble of initial radius $R_0$ is trapped between two glass plates, at distance $z_0$ from each other, with $z_0 << R_0$.
Hence, in the model, we consider an idealized setting with a cylindrical axially-symmetric bubble inside a liquid medium.
The liquid medium surrounding the bubble is considered as infinitely extended, thus neglecting finite size effects due to the boundaries. For the purpose of simulations, it is treated as water. This is justified, as the minimal medium is composed mainly by water with minimal amounts of salts, amino acids and lactate  \cite{mel14,sch16} and it is in agreement with previous findings on similar cultivation media \cite{adl12,men17}.
The air inside the bubble is treated as a binary mixture of perfect gases, nitrogen and oxygen, diffusing into the liquid causing the bubble to shrink in time. The oxygen is additionally depleted by the bacteria, while the nitrogen remains entirely passive. We assume that the bubble has a spatially uniform pressure and composition, as well as shape stability. The temperature of the whole system, $T_\infty$, is considered homogeneous and constant in time. The model consists of one ordinary differential equation (ODE), describing the time evolution of the radius of the bubble $R(t)$, and three partial differential equations (PDE), describing the time evolution of three concentration fields inside the liquid: the oxygen $c(r,t)$, the nitrogen $n(r,t)$, and the \textit{Shewanella} $s(r,t)$. Here $t$ is time and $r$ is the radial coordinate, measured from the centre of the bubble.

We derive the equation for the radial bubble dynamics as follows. The total gas pressure inside the bubble is $p_g = p_{O_2} + p_{N_2}$, with $p_{O_2}$ and $p_{N_2}$ the partial pressure of the oxygen
and the nitrogen respectively. In the cylindrical case \beq  p_g = p_{\infty} + \frac{\sigma}{R} \, , \label{Eq:total gas pressure}\eeq where $p_{\infty}$ is the pressure of the liquid at infinity, $\sigma$ the surface tension and $\sigma/R$ the Laplace pressure. We differentiate \Eq{total gas pressure} with respect to time, obtaining \beq \dot{p}_g = -\sigma \frac{\dot{R}}{R^2} \, ,\label{Eq:dp/dt gas, 1} \eeq where the dots denote time derivatives.
For each gas, the perfect gas law holds
\beq p_{O_2}  = \rho_{O_2} \mathcal{R}^*_{O_2} T_{\infty} \label{Eq:perfect gas law, O2} \, , \eeq
\beq p_{N_2}  = \rho_{N_2} \mathcal{R}^*_{N_2} T_{\infty} \label{Eq:perfect gas law, N2} \, , \eeq
with $\mathcal{R}^*_{O_2}$, $\mathcal{R}^*_{N_2}$ the specific gas constants and $\rho_{O_2}$, $\rho_{N_2}$ the densities of oxygen and nitrogen inside the bubble, respectively. We express such densities as \mbox{$\rho_{O_2} = m_{O_2} / (\pi R^2 z_0)$}  and \mbox{$\rho_{N_2} = m_{N_2} / (\pi R^2z_0)$}, where $m_{O_2}$, $m_{N_2}$ are the masses of oxygen and nitrogen inside the bubble. By adding up \Eq{perfect gas law, O2} and \Eq{perfect gas law, N2}, side by side, and deriving the resulting equation respect to time, we obtain \beq \dot{p}_g  = -2 \frac{\dot{R}}{R} p_g + \frac{T_{\infty}}{\pi R^2 z_0}(\mathcal{R}^*_{O_2} \dot{m}_{O_2} + \mathcal{R}^*_{N_2} \dot{m}_{N_2}) \, , \label{Eq:dp/dt gas, 2}\eeq where $\dot{m}_{O_2}$, $\dot{m}_{N_2}$ are the diffusive mass fluxes of oxygen and nitrogen, respectively. We assume that the diffusive process takes place only through the lateral area of the bubble, $2\pi R z_0$, and not through the top and bottom areas (in the experiment, in contact with the glass plates). Thus, \beq \dot{m}_{O_2} = -2\pi R z_0 D_{O_2} \pder{c}{r}\biggr|_{R(t)} \, , \label{Eq:diffusive fluxes, O2}\eeq
\beq \dot{m}_{N_2} = -2\pi R z_0 D_{N_2} \pder{n}{r}\biggr|_{R(t)}  \, , \label{Eq:diffusive fluxes, N2}  \eeq
with $D_{O_2}$, $D_{N_2}$ the diffusive constants of the oxygen and the nitrogen, respectively. By equating \Eq{dp/dt gas, 1} and \Eq{dp/dt gas, 2}, we derive the equation for the radial evolution in time \beq \dot{R} = \frac{T_{\infty}}{ p_g - \frac{\sigma}{2R} } \biggl[ D_{O_2} \mathcal{R}^*_{O_2} \pder{c}{r}\biggr|_{R(t)} + D_{N_2} \mathcal{R}^*_{O_2} \pder{n}{r} \biggr|_{R(t)} \biggr] \, . \label{Eq:dR/dt}\eeq

The evolution of the \emph{Shewanella} concentration field $s(r,t)$ is described by
\beq \pder{s}{t} + \frac{R\dot{R}}{r}\pder{s}{r} = - \nabla \cdot J +  Q s \, , \label{Eq:Shewa concentrat eq, 1} \eeq
where $\frac{R\dot{R}}{r} \pder{s}{r}$ is the convective term, due to the fact that the bubble is shrinking dragging along the \emph{Shewanella} \cite{shp13b}, $Q$ is the specific growth rate of the bacteria and $J$ is the flux, combining the bacterial diffusion, due to their random motility, and the chemotaxis along the gradient of the chemoattractant, \ie the oxygen \cite{li11}. We express such a flux as
\beq J = -\mu(c) \nabla s + V_s s  \, , \label{Eq:Shewa flux} \eeq
where $\mu(c)$ is the random motility coefficient of \emph{Shewanella} and $V_s = V_s(c, \nabla c)$ is their chemotactic velocity.
To account for the transition between aerobic/anaerobic functioning, we take $\mu(c) = \mu_0 \text{H}(c - c_T)$ with $\mu_0$  a constant value and $\text{H}(c-c_T)$ the Heaviside function centered around $c_T$, the oxygen concentration below which MR-1 turns anaerobic.
The chemotactic \mbox{velocity} $V_s$ is expressed as
\beq
V_s = \chi_0 \alpha(c) \nabla c  \, ,
\label{Eq:chemotactic velocity}
\eeq
where $\chi_0$ is the constant chemotactic sensitivity. The function $\alpha(c)$ models the relationship between the intensity of the chemotactic response and the local oxygen concentration. We adopt the heuristic formulation
\beq
\alpha(c) = \frac{K_D}{(K_D + c)^2} \text{H}(c-c_T) [1 - \text{H}(c-c_{opt})] \, ,
\label{Eq:alpha(c) heuristic formulation}
\eeq
where $K_D$ is the receptor dissociation constant and $c_{opt}$ is the optimal oxygen concentration for the MR-1. 
In the numerical treatment, the Heaviside functions are replaced by their smooth version $\text{H}_{\Delta c_T}(c-c_T)$ and  $\text{H}_{\Delta c_{opt}}(c-c_{opt})$, smeared over the intervals $\Delta c_T$ and $\Delta c_{opt}$ respectively. The smeared Heaviside function $\text{H}_{\epsilon}(x)$ on the domain $x$ is defined as \cite{xu12}
\beq
 \text{H}_\epsilon(x) =
\begin{cases}
 0,                                                                                                               & \mbox{ if } x < -\epsilon   \\
 \frac{1}{2} \biggl[1 + \frac{x}{\epsilon} + \frac{1}{\pi} \text{sin}\biggl(\pi \frac{x}{\epsilon}\biggr)\biggr], & \mbox{ if }|x|\leq \epsilon \\
 1, & \mbox{ else }\, .
\end{cases}
\label{Eq:smeared Heaviside fn}\eeq
The first factor at the right hand side of \Eq{alpha(c) heuristic formulation} reproduces the standard formulation of the chemotactic response \cite{lap76,riv89}, accounting for down-regulation (the effect of attractant concentration on the number of expressed cell surface receptors) and receptor saturation, while the last two factors take into account the suppression of aerotaxis when the bacteria reach their favourite concentration range and the transition to anaerobic functioning at low oxygen concentration (see Fig.~S7). By neglecting these additional effects, we could still reproduce the position of the peaks in the aerotactic bands observed in the experiments, but not the shape of the curves, yielding to a global error of $\sim60-70\%$ on the density profiles. A similar error was introduced by using other commonly adopted expressions for the chemotactic velocity, such as $V_s = \frac{2}{3} V \text{tanh} \biggl[\frac{\chi_0}{2V} \frac{\nabla c}{(K_d+c)^2}\biggr]$ with $V$ the swimming speed of the bacteria \cite{riv89}, and $V_s = \chi_0 \frac{\nabla c}{(K_1+c)(K_2+c)}$, with $K_1$ and $K_2$ two constants \cite{men17}. \\
We assume that the cell growth rate is limited by the oxygen concentration and we express it according to the Monod model \cite{li11,tan07} as
\beq \label{Eq:Shewa growth}  Q = \nu \frac{c}{C_s + c}\, ,\eeq
where $\nu$ is the maximum specific growth rate of the bacteria and $C_s$ is the half-saturation constant for the oxygen consumption.
Upon substitution of Eqs.(\ref{Eq:Shewa flux}), (\ref{Eq:chemotactic velocity}) and (\ref{Eq:Shewa growth}) inside \Eq{Shewa concentrat eq, 1}, we obtain
\beq \pder{s}{t} + \frac{R\dot{R}}{r}\pder{s}{r} = \nabla \cdot [\mu(c) \nabla s - \chi_0 \alpha(c) \nabla c  s] +  \nu \frac{c}{C_s + c} \, . \label{Eq:Shewa concentrat eq, 2} \eeq

The nitrogen concentration field $n(r,t)$ evolves following the
standard advection-diffusion equation \cite{str11}
\beq \label{Eq:gas concentrat eq, N2} \pder{n}{t}+
\frac{R\dot{R}}{r}\pder{n}{r}= D_{N_2} \nabla^2 n \, . \eeq

The oxygen concentration inside the liquid, $c(r,t)$ is described by an advection-diffusion equation with the addition of a consumption term
\beq \label{Eq:gas concentrat eq, O2} \pder{c}{t}+
\frac{R\dot{R}}{r}\pder{c}{r}= D_{O_2} \nabla^2c-A_0s\frac{c}{C_s+c}\, , \eeq
where $A_0$ is the maximum specific consumption rate of the oxygen by the bacteria and can be expressed as $A_0 = \nu Y$, with $Y$ the \textit{Shewanella} yield coefficient for the oxygen. Therefore, the last term at the right-hand-side of \Eq{gas concentrat eq, O2} represents the consumption of the oxygen by the bacteria.

The boundary conditions are defined as follows. Far from the bubble, at infinite, we assume a flat profile for all the concentration fields; hence,\beq \pder{c}{r}(r\rightarrow\infty,t)=0   \label{Eq:cdt infty, O2}\, , \eeq
\beq \pder{n}{r}(r\rightarrow\infty,t)=0   \label{Eq:cdt infty, N2}\, , \eeq
\beq \pder{s}{r}(r\rightarrow\infty,t)=0   \label{Eq:cdt infty, Shewa}\, . \eeq
At the bubble wall, the concentration of oxygen and nitrogen dissolved in the liquid, $c|_R$ and $n|_R$, are related to their partial pressure inside the bubble, $p_{N_2}$ and $p_{O_2}$, by Henry's law:
\beq \label{Eq:Henry's law, O2} p_{O_2} = k_{O_2} c|_{R(t)} \, , \eeq
\beq \label{Eq:Henry's law, N2} p_{N_2} = k_{N_2} n|_{R(t)} \, ,\eeq
where $k_{O_2}$ and $k_{N_2}$ are the Henry's constants for oxygen and nitrogen, respectively.
For the \emph{Shewanella}, we account for the oxygen-regulated partial adsorption mechanism at an air-liquid interface, described in 'Results. The oxygen concentration regulates the biofilm formation', by imposing the bacterial flux $J$ at the bubble wall
\beq
J|_{R(t)} = r_as\epsilon(c)|_{R(t)} \, ,
\label{Eq:BC Shewanella at bubble wall R(t)}
\eeq
where $r_a$ is the adsorption constant and $\epsilon(c)$ an increasing function of the oxygen concentration. Different functions $\epsilon(c)$ can be adopted, to describe the adsorption mechanism with different degrees of precision. However, such an investigation lies beyond the scope of the present work; hence, we take for simplicity $\epsilon(c) = c$.

\subsection*{Numerical method}
\label{sec:Numerical method}
For the solution of the diffusion equations \mbox{Eqs.~(\ref{Eq:Shewa concentrat eq, 2})-(\ref{Eq:gas concentrat eq, O2})}, we adopt the pseudo-spectral collocation method described in Ref. \cite{str11}. The reader is referred to that
reference for details. We perform the coordinate transformation
\beq \label{Eq:change_var2} \frac{1}{\xi}=1+\frac{r/R(t)-1}{l} \,
, \eeq in order to map the semi-infinite range $R(t)\leq r <
\infty$ into the finite range $1 \geq \xi \geq 0$. In
(\ref{Eq:change_var2}), the constant $l$ is chosen based on the distance $lR_0$, over which
we expect the phenomenon to take place. From the analysis of the experimental results, we take $l = 10$.

We reformulate the equations describing the evolution of the concentration fields,
by substituting the new variable $\xi$ inside Eqs.(\ref{Eq:Shewa concentrat eq, 2})-(\ref{Eq:gas concentrat eq, O2}), and we
expand the \textit{Shewanella} and gas concentration fields
in truncated Chebyshev series: \beq
 \label{Eq:Shewa approx}
s(t,\xi) \approx\sum_{k=0}^{N}a_k(t)T_{2k}(\xi)\,,
\eeq \beq \label{Eq:c approx}
c(t,\xi) \approx\sum_{k=0}^{N}b_k(t)T_{2k}(\xi)\,,
\eeq \beq \label{Eq:n approx}
n(t,\xi) \approx\sum_{k=0}^{N}c_k(t)T_{2k}(\xi)\,.
\eeq Here, $T_{2k}(\xi)$ are the even Chebyshev polynomials and $a_k(t)$, $b_k(t)$, $c_k(t)$ are the new unknowns of the system, function of time.
The coupled equations for the $a_k(t)$, $b_k(t)$, $c_k(t)$,
arising upon substitution of Eqs.(\ref{Eq:Shewa approx})-(\ref{Eq:n
approx}) into Eqs.(\ref{Eq:Shewa concentrat
eq, 2})-(\ref{Eq:gas concentrat
eq, O2}), are solved by collocation \cite{str11}. Convergence is reached by using
45 terms in the Chebyshev expansions, with 45 Gauss-Lobatto
collocation points. In the limiting case where no bacteria are present (pure gas diffusion), we
verify that the code developed for this work gives the same results as those used in \mbox{Refs.~\cite{str12_thesis,shp13b}}, which had been validated both against analytical solutions
and experimental findings.


\subsection*{Estimate of parameters}
\label{sec:Estimate of parameters}

The half-saturation constant for the oxygen consumption \mbox{$C_s = 6 \cdot 10^{-3}$ mM} and the \textit{Shewanella} yield coefficient for the oxygen \mbox{$Y$ = 27 g$_{\text{dry cells}}$ / mol$_{\textsc{O}_2}$} are taken from the literature \cite{tan07}, with the mass of a single bacterium equal to \mbox{$4.64 \cdot 10^{-13}$ g} \cite{ren13}. 

The random motility coefficient $\mu_0$ is derived from the plot of the mean square displacement \mbox{\emph{MSD}} versus time, in the experiment with uniform oxygen concentration (closed setup, no bubble). The \mbox{\emph{MSD}} is calculated from the bacterial trajectories as \mbox{\emph{MSD}$(t) = <[\textbf{y}_i(t+t_0) - \textbf{y}_i(t_0)]^2>_{i,t_0}$}, where $t_0$ is an arbitrary initial time, $\textbf{y}_i(t+t_0)$ is the position of the $i^{th}$ bacterium at time $t+t_0$, and $<\cdot>_{i,t_0}$ denotes the average over all bacteria and initial times. By fitting the resulting graph with the formula~\mbox{\emph{MSD} = $4\mu_0 t$}, we find \mbox{$\mu_0$ = 24 {\micro}m$^2$/s}, in agreement with the values previously reported for \mbox{MR-1} \cite{li11}.

The maximum specific growth rate of the bacteria $\nu$ is derived from the experimental growth curve of \mbox{MR-1} in fully aerobic well mixed conditions (open flask with 150 rpm shaking), in the following way. The growth of \mbox{MR-1} is described by the equation: \mbox{$\dot{s} = Qs$}. We assume that the oxygen concentration is constant and equal to the saturation concentration at ambient pressure \mbox{$c_0$ = 0.278 mM}. Hence, the time integration gives \mbox{$s(t)=s_0 \text{exp}^{Q(t-t_0)}$}, with $t_0$ the initial time and $s_0$ the initial bacterial concentration. From the slope of the $\textmd{ln}(s)$ versus time, in the exponential growth phase of the bacteria, we estimate $Q$. From $Q$, using \Eq{Shewa growth}, we find \mbox{$\nu$ = 0.29 h$^{-1}$}. Such a value is in perfect agreement with the theoretical one of \mbox{$0.3$ h$^{-1}$}, predicted by the formula \mbox{$\nu = \nu_L^{max} c_L / K_L$}  \cite{bla97_BOOK}, where \mbox{$\nu_L^{max}$ = 0.47 h$^{-1}$} is the maximum specific growth rate of \mbox{MR-1} in fully aerobic conditions on lactate, \mbox{$c_L$ = 24 mM} is the lactate concentration in the minimal medium, and \mbox{$K_L$ = 13.2 mM} is the Monod concentration constant for lactate \cite{tan07}.

The oxygen concentration \mbox{$c_T = 9.37 \cdot 10^{-3}$ mM}, at which \textit{S. oneidensis} switches from aerobic to anaerobic functioning, is known from the literature \cite{mor16}. We observe in the experiments that the aerobic/anaerobic transition is a sharp one, with the bacteria stopping in an abrupt fashion. Hence, the range of oxygen concentrations over which the transition takes place, $\Delta c_T$, should be small. We verify that the bacterial profiles from simulations do not change significantly, when varying $\Delta c_T$ within the range $\Delta c_T / c_T \in[\frac{1}{6},1]$.

To estimate the preferred oxygen concentration of \mbox{MR-1}, $c_{opt}$, we adopt the following procedure. We consider the steady state bacterial profile in the open setup (Fig.~S2.a), encompassing a non-moving aerotactic band. The aerotactic band has a maximum, located at a fixed distance \mbox{$L_b \sim$ 1170 {\micro}m} from the air-liquid interface. We take the oxygen concentration in $L_b$ as the optimal concentration for the bacteria, $c_{opt}$.
To derive such a value, we calculate the steady state oxygen concentration profile, by numerically integrating the equation for the oxygen concentration
\beq
D_{O_2} \nabla^2 c = A_0 s^* \frac{c}{C_s + c} \, . \label{Eq: O2 steady state eqn}
\eeq Here $s^*(x)$ is the experimental steady state profile of the bacterial density, and $x$ is the distance from the interface. For the integration we use the Matlab ODE solver \textit{ode45}. To avoid the uncertainties due to the scattered light at the air-liquid interface, we start the integration from the far field. In particular, we start at \mbox{$L_{\infty} \sim$ 1460 {\micro}m}, the first point where the bacteria do not move, immediately beyond the band's edge. We take \mbox{$c\big|_{L_\infty}=c_T$} and \mbox{$\pder{c}{x}\big|_{L_\infty} = 0$} as boundary conditions. From the calculated steady state oxygen profile (Fig.~S2.b), we find \mbox{$c_{opt} = 2.4 \cdot 10^{-2}$ mM}. The values of the oxygen concentration at the two edges of the band, located at $L_1$ and $L_2$, represent the favourite oxygen concentration range of the MR-1. For \mbox{$L_1$ = 1080 {\micro}m} and \mbox{$L_2$ = 1250 {\micro}m}, we find the oxygen concentration values \mbox{$c_1 = 3.1\cdot 10^{-2}$ mM} and \mbox{$c_2 = 1.6\cdot 10^{-2}$ mM}, respectively. We take this range as representative of $\Delta c_{opt}$. Such values are compatible with the preferred oxygen concentrations of other aerotactic bacteria \cite{joh97}.

The adsorption constant $r_a$, the receptor dissociation constant $K_D$ and the chemotactic sensitivity $\chi_0$ are taken as fitting parameters. In particular, $r_a$ is used to fit the bacterial concentration and the slope of the bacterial profile at the bubble wall. The parameters $\chi_0$ and $K_D$ determine the height and the position of the peak in the band. Over a sample of 10 different experiments, we find the optimal values of \mbox{$\chi_0 = 1.56 \cdot 10^3$ {\micro}m$^2$/s} (corresponding to \mbox{$\chi_0/\mu_0 = 65$}), \mbox{$r_a  = 9.4\cdot 10^{-2}\,(\text{mM}\cdot \text{h})^{-1}$} and \mbox{$K_D = 3.12$ mM}. The latter has the same order of magnitude as the values reported in the literature for the receptor dissociation constant of \textit{Shewanella} for fumarate \cite{li11}. 

\clearpage



\clearpage

\begin{figure}
\centering
\includegraphics[width=0.8\textwidth]{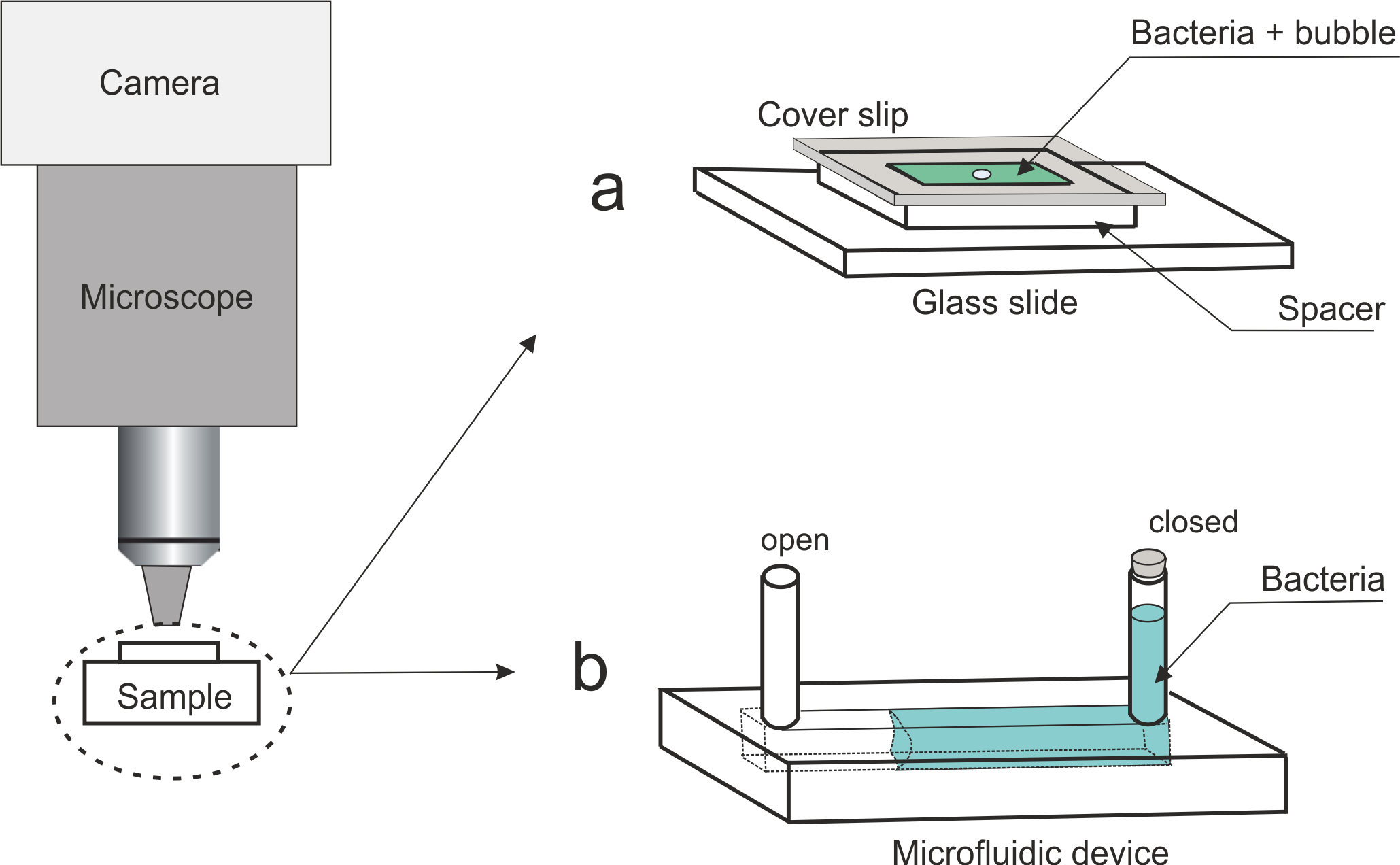}
\caption{\textbf{Experimental setup.}(a) Closed setup. (b) Open setup.}
 \label{Fig:Experimental setup}
\end{figure}

\begin{figure}
\centering
\includegraphics[width=0.7\textwidth]{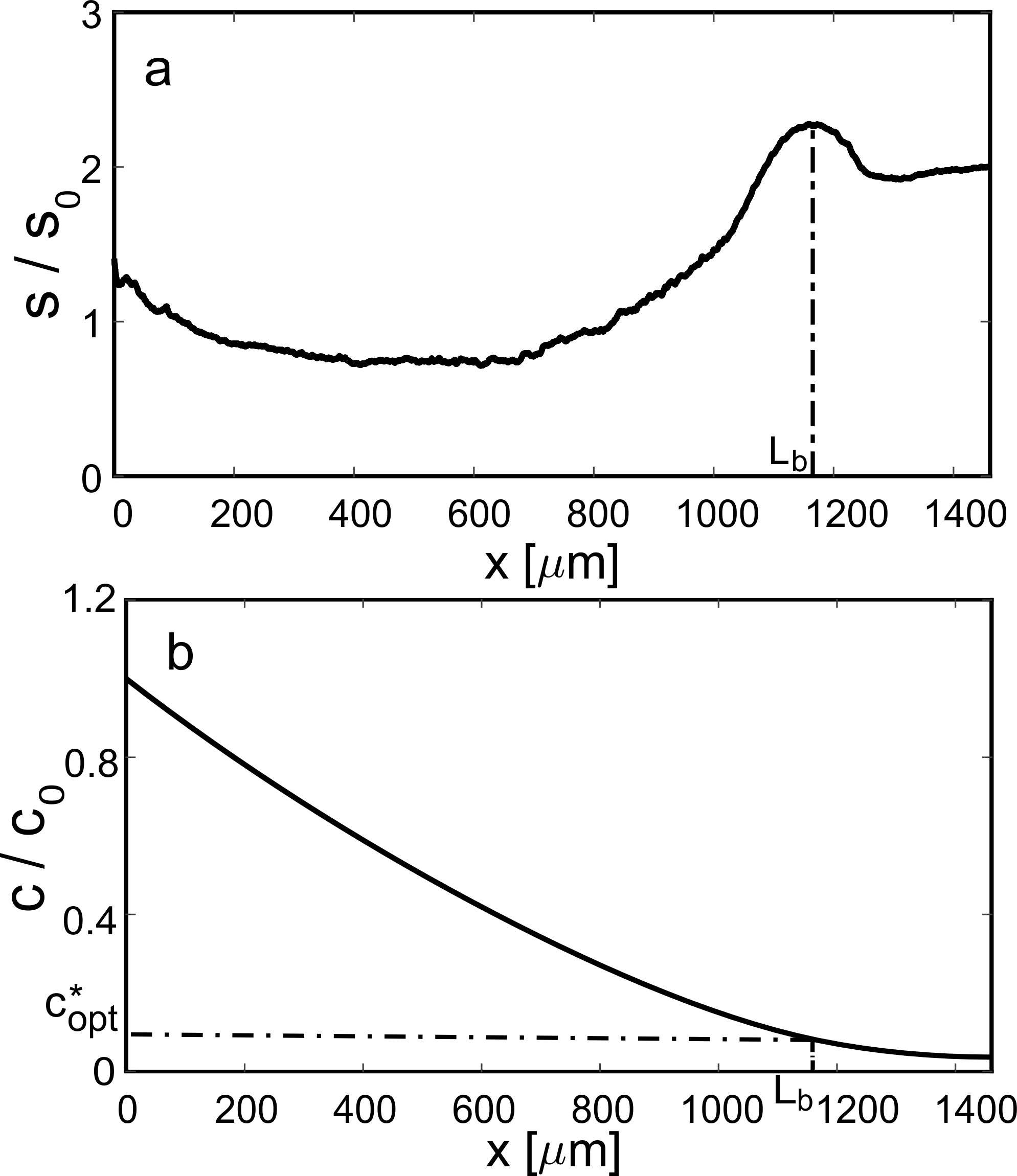}
\caption{\textbf{Steady state profiles of bacterial and oxygen concentrations, in the open setup.} (a) Bacterial concentration, measured from experiments in the open setup, 120~minutes after preparation of the sample; $x$ is the distance from the air-liquid interface and \mbox{$s_0 = 2.9 \cdot 10^7$} the initial bacterial concentration. (b) Oxygen concentration profile, numerically derived from (a) following the procedure described in Sec.~'Estimate of parameters'; \mbox{$c_0 = 0.275$ mM} is the oxygen concentration at the interface, $L_b$ the position of the maximum in the band and \mbox{$c_{opt}^* = c_{opt}/c_0$} the optimal oxygen concentration for \mbox{MR-1}.}
\label{Fig:Steady state}
\end{figure}

\begin{figure}
\centering
\includegraphics[width=0.7\textwidth]{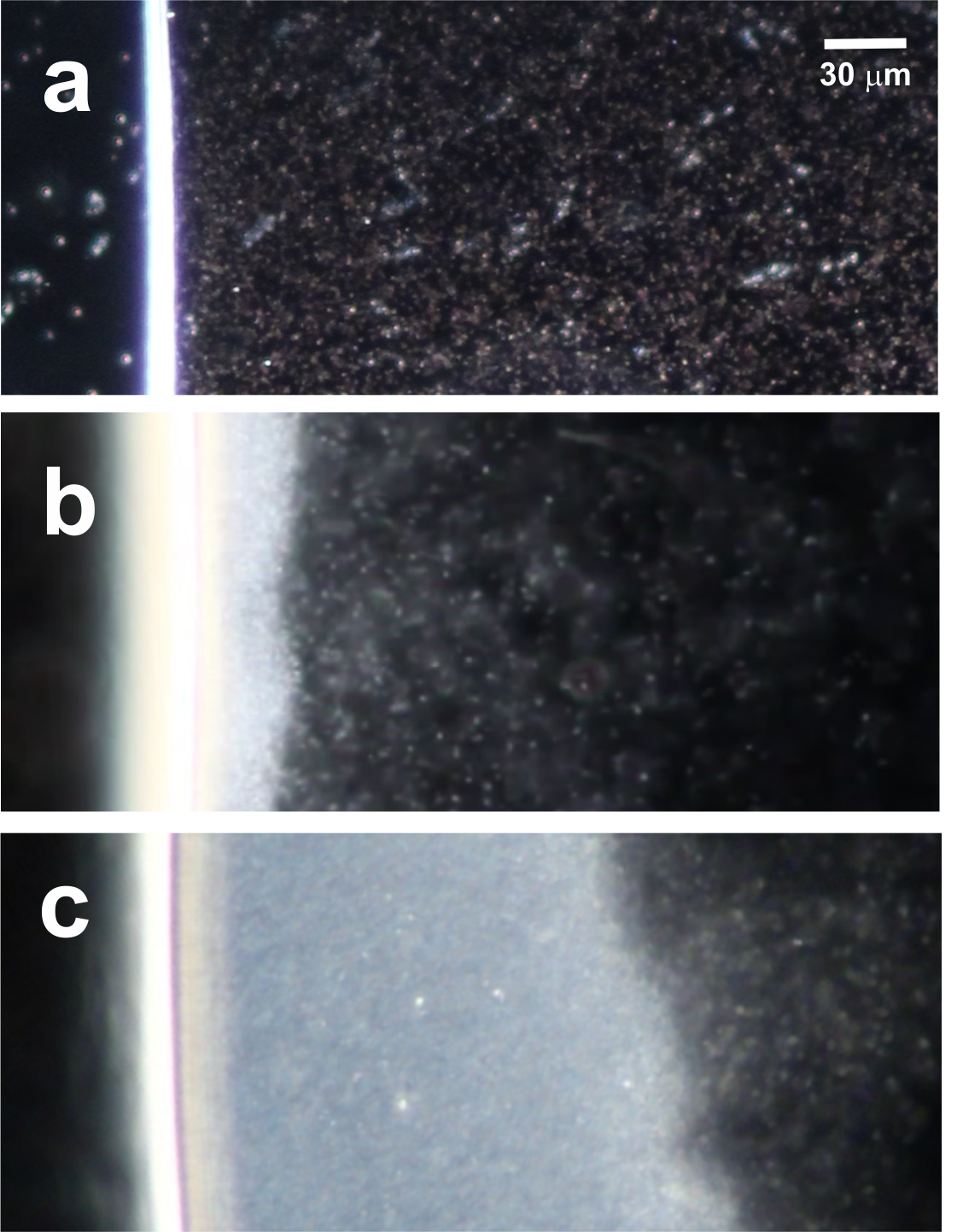}
\caption{\textbf{Time evolution of the biofilm precursor with unlimited oxygen supply.} Imaging is taken (a) 5~min (b) 90~min (c) 180~min after preparing the sample.}
\label{Fig:Biofilm open air}
\end{figure}

\begin{figure}
\centering
\includegraphics[width=0.7\textwidth]{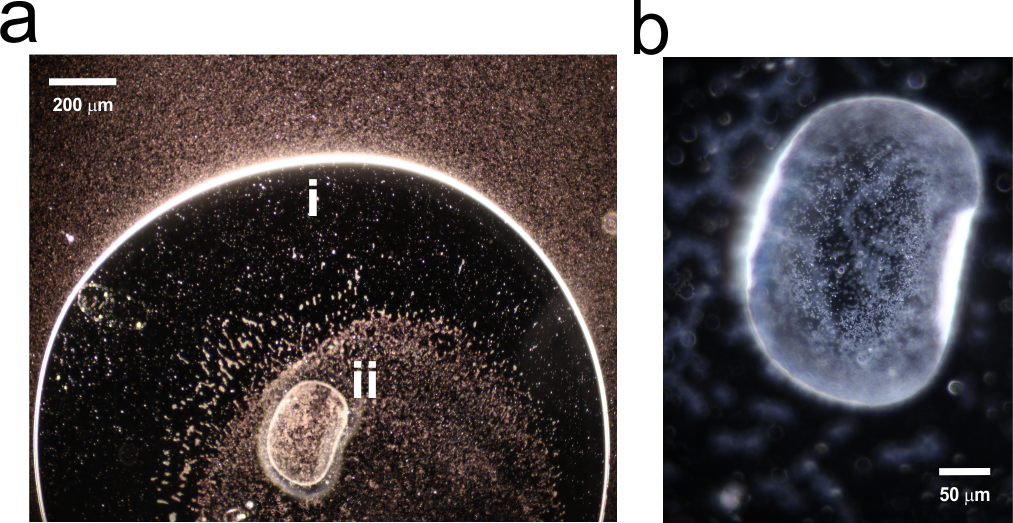}
\caption{\textbf{Biofilm formation at the air-liquid interface.} (a) Droplet~(ii) captured inside a bubble~(i) in the closed setup. A moving aerotactic band surrounds the bubble (not displayed in the figure). (b) Zoom-in on the droplet displays the pellicle formation process.}
\label{Fig:Droplet inside}
\end{figure}


\begin{figure}
\centering
\includegraphics[width=0.7\textwidth]{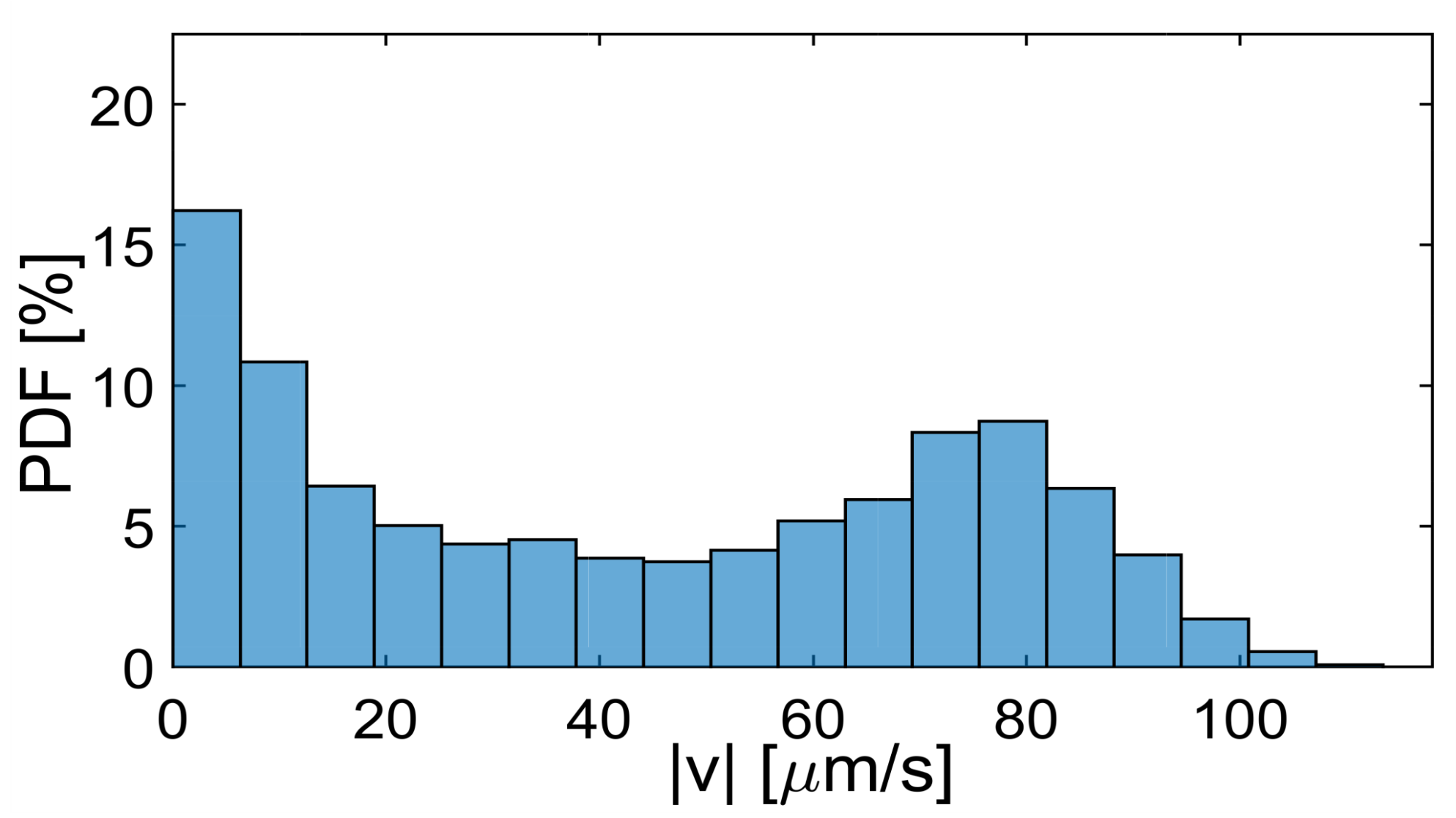}
\caption{\textbf{PDF of instantaneous velocities of all motile bacteria, including stop-and-go.} The pauses performed by motile bacteria along their trajectories are kept into account. The closed setup without bubble is imaged 1~minute after sealing the sample. The figure includes 4,600~points.}
 \label{Fig:PDF instantaneous velocities, no removed}
\end{figure}

\begin{figure}
\centering
\includegraphics[width=0.7\textwidth]{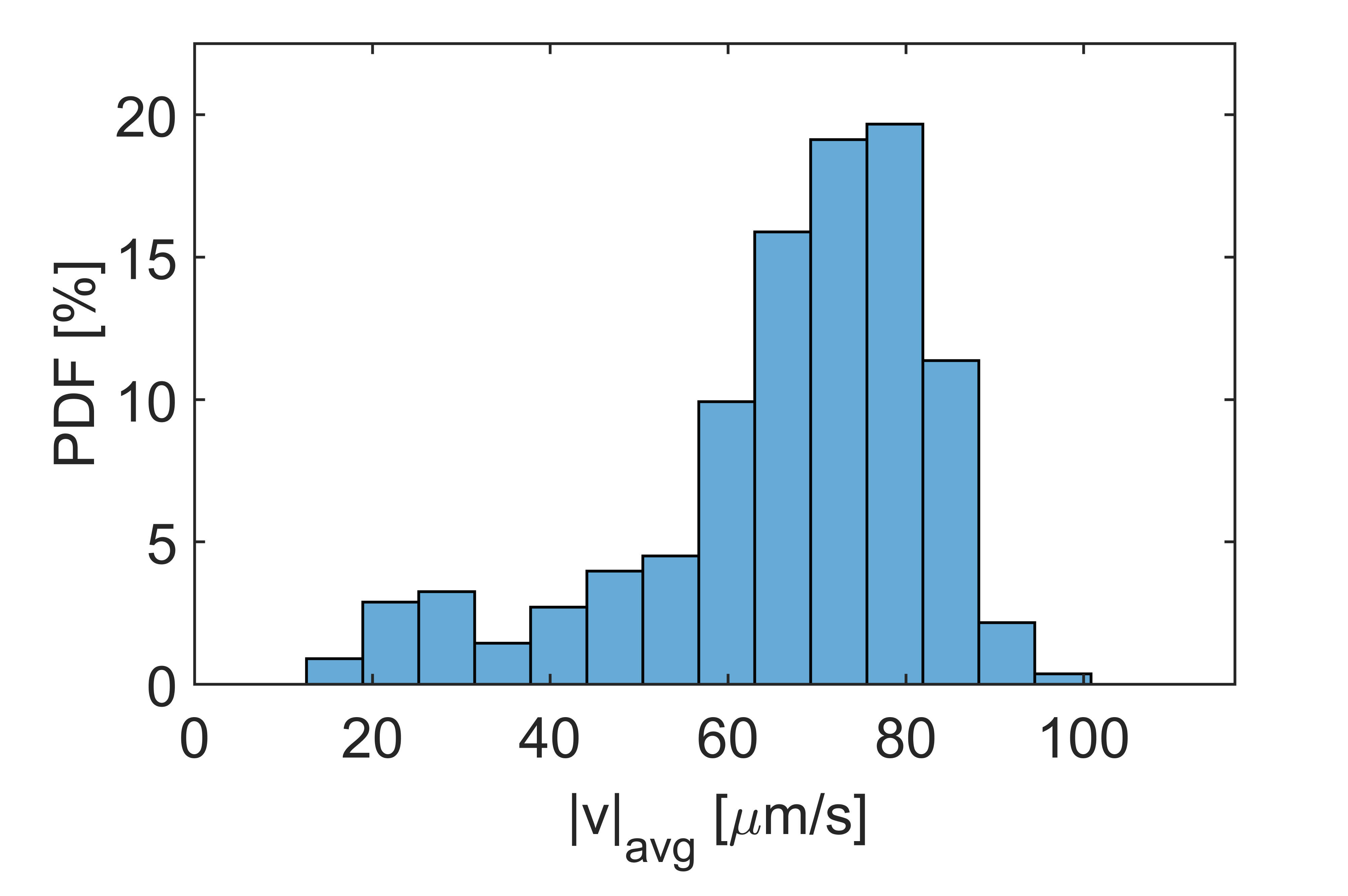}
\caption{\textbf{PDF of average velocities along the bacterial trajectories.} The closed setup without bubble is imaged 1~minute after closing the sample. The figure includes $\sim$~700~trajectories.}
 \label{Fig:PDF velocities, instantaneous and average}
\end{figure}

\begin{figure}
\centering
\includegraphics[width=0.7\textwidth]{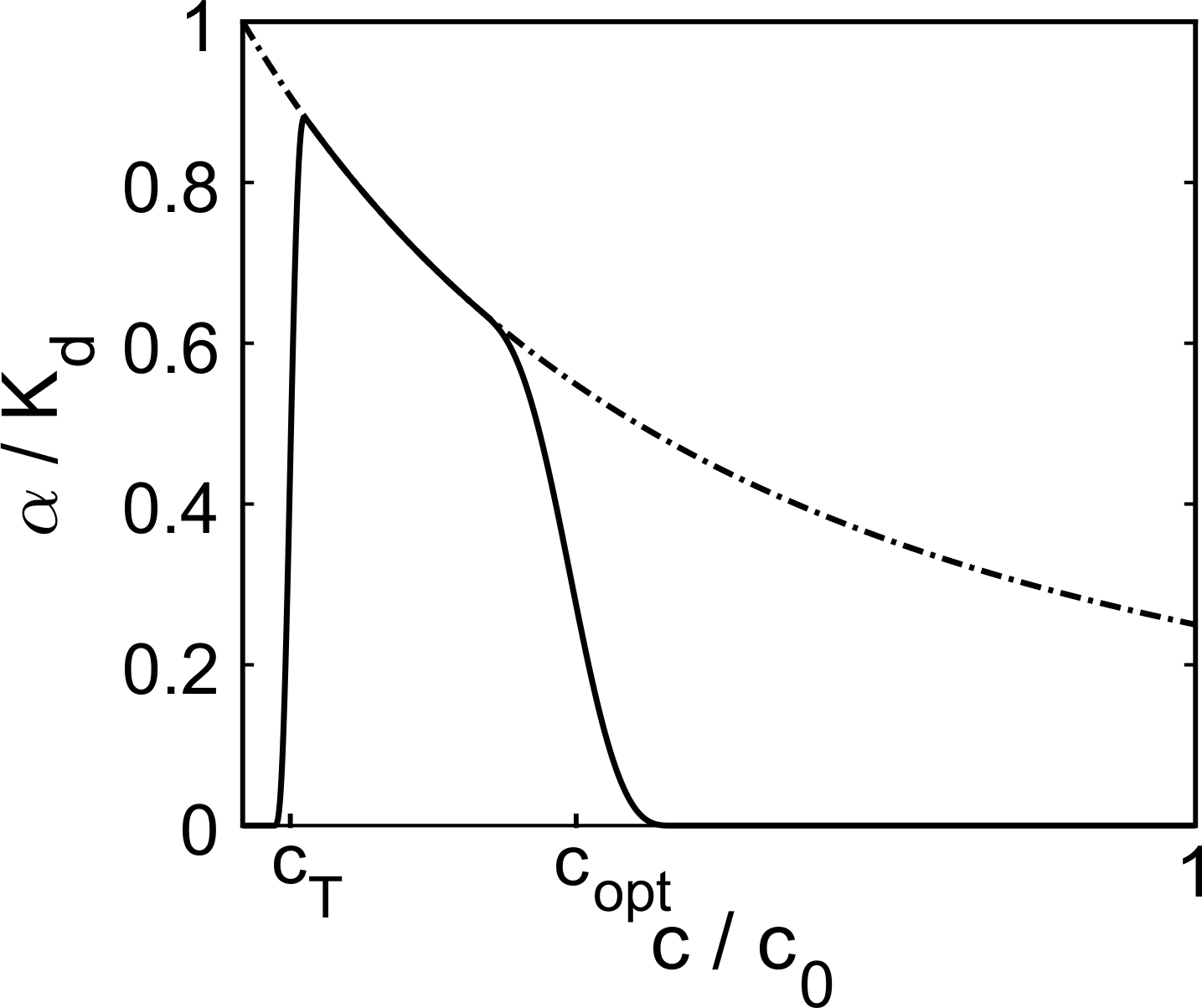}
\caption{\textbf{Dependence of the chemotactic response $\alpha$ on the local oxygen concentration $c$.} Comparison between the function $\alpha(c)$ used in Ref.~\cite{lap76} (dash-dotted line) and our heuristic formulation Eq.~(5) (solid line).}
\label{Fig:Comparison alpha(c)}
\end{figure}

\clearpage
\begin{video}
\textbf{Evolution of the aerotactic band around a confined air bubble.} The initial bubble radius is \mbox{$R_0 = 514$ {\micro}m}. The white circumference is the bubble, the surrounding halo is the band. The closed setup is imaged 20~minutes after sealing the sample. The video is displayed 120~times faster than natural time.
\end{video}

\begin{video}
\textbf{Bacterial motility in the aerotactic band surrounding a confined air bubble.} The closed setup is imaged 35~minutes after sealing the sample. The thick white line at the bottom left corner is the bubble wall; moving outwards in the radial direction, one can distinguish: an area with higher luminosity (\ie higher bacterial density) and motile bacteria (the aerotactic band), a thin area with reduced bacterial density (the depletion layer), an area with non-motile vibrating bacteria (bacteria in anaerobic functioning state). The video is displayed in natural time.
\end{video}

\begin{video}
\textbf{Biofilm formation at the air-liquid interface.} Surface of a droplet trapped inside a bubble, in the closed setup, at two different instants: 10~minutes after sealing the sample, the bacteria gather in progressively growing active clusters, forming the pellicle; 40~minutes after sealing the sample, when the oxygen has been depleted, the bacterial motility is suppressed and the pellicle formation stops. The video is displayed in natural time.
\end{video}

\begin{video}
\textbf{Active bacterial clusters forming the air-liquid biofilm.} Surface of a droplet trapped inside a bubble, in the closed setup, 10~minutes after sealing the sample. The video is displayed in natural time.
\end{video}

\begin{video}
\textbf{Stop-and-go type of motion in \mbox{MR-1}.} Example of trajectory where the bacterium alternates pauses and runs. The video is displayed 10~times slower than natural time.
\end{video}

\begin{video}
\textbf{Alternated sideways/longitudinal swimming in \mbox{MR-1}.} Two examples of trajectories where the bacteria alternate sideways and longitudinal swimming. The video is displayed 4~times slower than natural time.
\end{video}